\documentclass[a4paper,prx,reprint,superscriptaddress,twocolumns,notitlepage,floatfix]{revtex4-2}
\usepackage[utf8]{inputenc}
\usepackage[T1]{fontenc}
\usepackage{amsmath}
\usepackage{amssymb}
\usepackage{commath}
\usepackage{mathrsfs}

\usepackage{graphicx}
\usepackage{amsmath}
\usepackage{mathrsfs}  
\usepackage{braket}
\usepackage{amssymb} 
\usepackage{mathtools}
\usepackage{csquotes}
\usepackage{array}
\usepackage{booktabs}
\usepackage{multirow}
\usepackage{xcolor}
\usepackage{tcolorbox}
\usepackage{siunitx}
\usepackage{pifont}
\usepackage{microtype}

\RequirePackage{expl3,xparse}
\ExplSyntaxOn
\NewDocumentCommand{\hm}{m}
{
    \tl_set:Nx \l_hm_contents {\ensuremath{#1}}
    \regex_replace_all:nnN { -(.) } { \c{overline}\cB\{ \1 \cE\} } \l_hm_contents
    \regex_replace_all:nnN { (.)\_(.) } { \1\c{c_math_subscript_token}\2 } \l_hm_contents
    \regex_replace_all:nnN { ([a-zA-Z]+) } { \c{text}\cB\{ \1 \cE\} } \l_hm_contents
    \tl_use:N \l_hm_contents
}

\NewDocumentCommand{\sch}{m}
{
    \tl_set:Nx \l_sch_contents {#1}
    \regex_replace_all:nnN { \_ }{} \l_sch_contents \regex_replace_all:nnN { (.)(.*) } { \1\c{c_math_subscript_token}\cB\{\2\cE\} } \l_sch_contents \regex_replace_all:nnN { (d|h|v|s) }{ \c{text}\cB\{\1\cE\} }  \l_sch_contents \regex_replace_all:nnN { \cS\  }{} \l_sch_contents \tl_set:Nn \l_sch_result {\ensuremath{\l_sch_contents}} \tl_use:N \l_sch_result
}

\NewDocumentCommand{\irrep}{m}
{
    \tl_set:Nx \l_sch_contents {#1}
    \regex_replace_all:nnN { \_ }{} \l_sch_contents \regex_replace_all:nnN { (.)(.*) } { \1\c{c_math_subscript_token}\cB\{\2\cE\} } \l_sch_contents \regex_replace_all:nnN { ([a-zA-Z]) }{ \c{text}\cB\{\1\cE\} }  \l_sch_contents \regex_replace_all:nnN { \cS\  }{} \l_sch_contents \tl_set:Nn \l_sch_result {\ensuremath{\l_sch_contents}} \tl_use:N \l_sch_result
}
\ExplSyntaxOff

\DeclareFontFamily{U}{crystallographicSymbol}{\hyphenchar\font=-1}
\DeclareFontShape{U}{crystallographicSymbol}{m}{n}{ <-> cryst}{}
 
\usepackage{hyperref}
\usepackage{bookmark}

\definecolor{tab_blue}{HTML}{1F77B4}
\definecolor{tab_orange}{HTML}{FF7F0E}
\definecolor{tab_green}{HTML}{2CA02C}
\definecolor{tab_red}{HTML}{D62728}
\definecolor{tab_purple}{HTML}{9467BD}
\definecolor{tab_brown}{HTML}{8C564B}
\definecolor{tab_pink}{HTML}{E377C2}
\definecolor{tab_gray}{HTML}{7F7F7F}
\definecolor{tab_olive}{HTML}{BCBD22}
\definecolor{tab_cyan}{HTML}{17BECF}

\hypersetup{breaklinks=true,
  colorlinks=true,
  linkcolor=tab_blue,
  filecolor=tab_blue,
  urlcolor=tab_blue,
  citecolor=tab_blue,
}

\def\Id{\text{Id}}

\def\diag{\ensuremath{\operatorname{diag}}}
\def\ee{{\rm e}}
\def\ii{{\rm i}}

\def\RR{\ensuremath{\mathbb{R}}}
\def\EE{\ensuremath{\mathbb{E}}}
\def\ZZ{\ensuremath{\mathbb{Z}}}
\def\CC{\ensuremath{\mathbb{C}}}

\AtBeginDocument{
\heavyrulewidth=.08em
\lightrulewidth=.05em
\cmidrulewidth=.03em
\belowrulesep=.65ex
\belowbottomsep=0pt
\aboverulesep=.4ex
\abovetopsep=0pt
\cmidrulesep=\doublerulesep
\cmidrulekern=.5em
\defaultaddspace=.5em
}

\tcbuselibrary{skins}
\tcbuselibrary{breakable}
\definecolor{boxhl}{HTML}{CEE0ED}
\definecolor{boxbg}{HTML}{FCFCFE}
\definecolor{otherbg}{HTML}{DEE6EA}
\definecolor{boxtc}{HTML}{33658A}
\newtcolorbox[auto counter]{example}[2][]{breakable,arc=0mm,sharp corners=all,
before upper={\parindent1em},
colback=boxbg,colframe=boxhl,coltitle=boxtc,fonttitle=\bfseries,title=Box~\thetcbcounter{} -- #2,#1}

\makeatletter\begin{document}

\title{Systematic generation of Hamiltonian families with dualities}
\author{Michel Fruchart}
\email{fruchart@uchicago.edu}
\affiliation{James Franck Institute, The University of Chicago, Chicago, IL 60637, USA}
\affiliation{Department of Physics, The University of Chicago, Chicago, IL 60637, USA}

\author{Claudia Yao}
\affiliation{James Franck Institute, The University of Chicago, Chicago, IL 60637, USA}
\affiliation{Department of Physics, The University of Chicago, Chicago, IL 60637, USA}

\author{Vincenzo Vitelli}
\email{vitelli@uchicago.edu}
\affiliation{James Franck Institute, The University of Chicago, Chicago, IL 60637, USA}
\affiliation{Department of Physics, The University of Chicago, Chicago, IL 60637, USA}
\affiliation{Kadanoff Center for Theoretical Physics, The University of Chicago, Chicago, IL 60637, USA}

\begin{abstract}
Dualities are hidden symmetries that map seemingly unrelated physical systems onto each other. The goal of this work is to systematically construct families of Hamiltonians endowed with a given duality and to provide a universal description of Hamiltonians families near self-dual points. 
We focus on tight-binding models (also known as coupled-mode theories), which provide an effective description of systems composed of coupled harmonic oscillators across physical domains.
We start by considering  the general case in which group-theoretical arguments suffice to construct families of Hamiltonians with dualities by combining irreducible representations of the duality operation in parameter space and in operator space.
When additional constraints due to system specific features are present, a purely group theoretic approach is no longer sufficient. To overcome this complication, we reformulate the existence of a duality as a minimization problem which is amenable to standard optimization and numerical continuation algorithms.
Combined with existing procedures to physically implement coupled-resonator Hamiltonians, our approach enables on demand design of photonic, mechanical, thermal, or electronic metamaterials with dualities. \end{abstract}

\maketitle

\def\Cr{\mathcal{C}}
\def\VV{\mathcal{V}}
\def\GG{\mathcal{G}}
\def\EE{\mathbb{E}}
\def\HH{\mathcal{H}}
\def\RR{\mathbb{R}}
\def\Fun{\mathscr{F}}
\def\notemph#1{#1}

\section{Introduction}

Symmetry groups and their representations are the ingredients that scientists and engineers pour in their cauldron~\cite{Michel2001,Nye1985,Malgrange2014}.
Where alchemists sought to make only gold, present day researchers were more ambitious and built metamaterials: synthetic materials with properties nowhere found in nature, 
such as negative indices of refractions or bulk moduli~\cite{Bertoldi2017,Kadic2019,Soukoulis2011}.

But what is really a symmetry? 
If we approximate a material as a set of regularly arranged points in space (a crystal), there is an easy answer:
symmetries are all transformations of space such as translations, rotations, reflections (in general, isometries) that leave this set of points invariant.
These are \notemph{spatial symmetries} (or more generally, space-time symmetries).
To describe a material, we also need to attach to each point some internal degrees of freedom. 
These could represent, for instance, the amplitude of the electric field, the temperature, the displacement of the particles, or the state of a spin at this point.
Spatial symmetries transform these internal degrees of freedom in a definite way, depending on whether they are scalars, vectors, pseudo-vectors, etc.
In addition, the system might possess \notemph{internal symmetries} that transform the internal degrees of freedom at each point independently (but leave the system globally invariant).
As an example, the Ising model is invariant under the inversion of all spins.

In general, symmetries are \notemph{transformations that leave the system invariant}. 
This leaves the door open for symmetries that are neither spatial symmetries, nor internal symmetries. 
Can such \notemph{hidden symmetries} exist? If so, are they just coincidences?
To address these questions, it will be helpful to consider families of systems depending on continuous parameters, instead of one system at a time.
Accordingly, we will consider families of linear operators $H(p)$ smoothly depending on a parameter $p$. 
For convenience we will refer to $H(p)$ as a Hamiltonian. Let us however emphasize that (i) this operator might describe systems such as mechanical networks, photonic crystals, diffusive systems, Markov chains, etc. as well as mean-field quantum Hamiltonians, (ii) we will not necessarily assume that $H(p)$ is Hermitian.
The Hamiltonian generates the dynamics of the system under consideration, through an equation of motion such as $\ii \partial_t \psi = H \psi$ (or a variant, like $\partial_t^2 \psi = H \psi$).
Here, the vector $\psi$ describes the state of the system, such as the electromagnetic field in a photonic crystal, the displacements of masses in an elastic network, or the wave function in a quantum system.
In this context, we focus on transformations acting linearly on this state, namely invertible (usually unitary) matrices $U$ mapping $\psi$ to $U \psi$. 
Symmetries are transformations that commute with the Hamiltonian, i.e. such that $U H U^{-1} = H$.
Symmetries and their breaking play a crucial part in determining the properties of a material, including phase transitions~\cite{Toledano1987}, band structures, topological phases~\cite{Bradlyn2017}, and physical responses~\cite{Nye1985,Malgrange2014}.
For instance, piezoelectricity or optical activity can only exist in crystals that do not possess a center of inversion~\cite{Nye1985,Malgrange2014}.

A duality acts in the same way as a symmetry, but it also changes the parameters in a certain way: it maps a system to another system.
Formally, a duality is defined by the combination of a transformation $U$ (a matrix acting on the $\psi$) and a smooth function $f$ defined on the space of parameters $p$, such as $U H(p) U^{-1} = H(f(p))$.
For instance, the properties of a system under a magnetic field $B$ (playing the role of the parameter $p$) are usually different when the magnetic field is inverted to $-B$. 
Yet, the behaviors of both systems can often be deduced from one another, as exemplified by the Onsager-Casimir relations in non-equilibrium thermodynamics~\cite{DeGrootMazur}. 
In this example, the duality is simply embodied in the time-reversal operator $\Theta$ and the corresponding Hamiltonians $H(B)$ and $H(-B)$ are related by $H(-B) = \Theta H(B) \Theta^{-1}$.
But the duality operator has no reason to be a simple symmetry, such as a spatial or internal symmetry.

Similar notions of dualities were suggested in systems ranging from self-assembled systems~\cite{Lei2021} to quantum rotors~\cite{Guarneri2020}.
Rich consequences arise when a duality is combined with other constraints such as symmetries or conservation laws, ranging from 
degeneracies in the band structure~\cite{Browne1984,Hou2013,Fruchart2020} and topological states~\cite{Liu2021,Danawe2021}
to symmetries in the phonon response~\cite{Gonella2020}
and constraints on the stiffness tensor of an elastic medium~\cite{Fruchart2020b}.

\begin{figure}
	\hspace*{-15pt}\includegraphics[width=9.5cm]{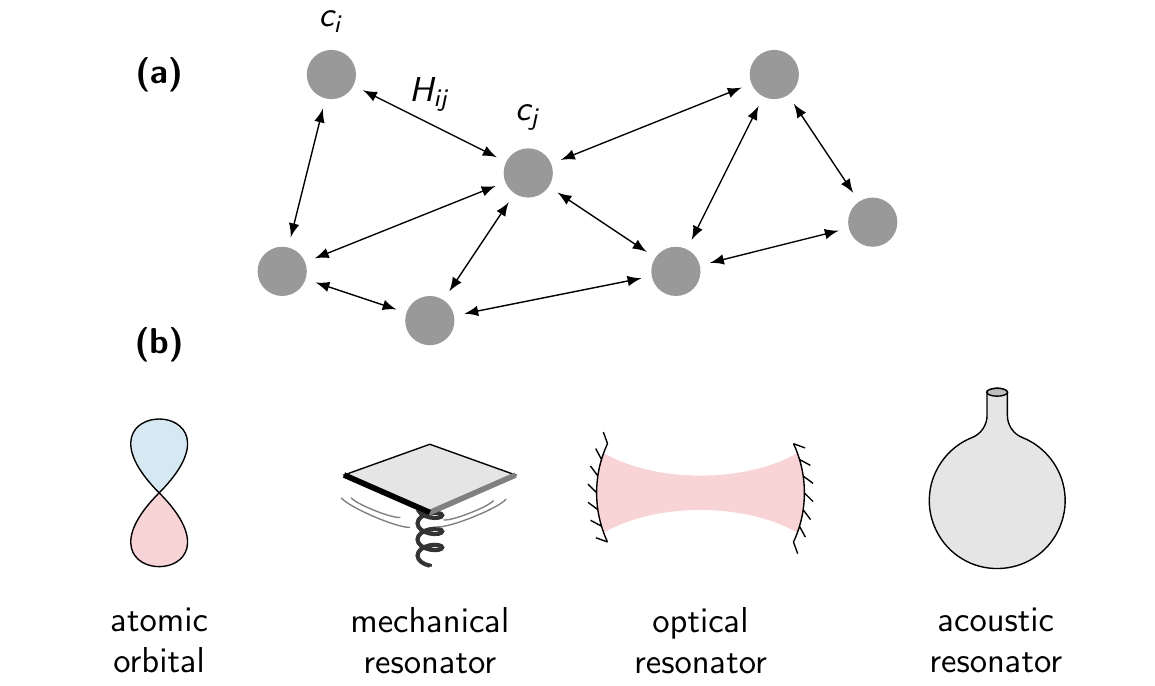}
    \caption{\label{figure_resonators_dualities}
    \textbf{Effective models.}
    We consider models consisting of coupled resonators $c_i$ (harmonic oscillators) coupled together with coupling constants $H_{i j}$.
    These can describe, effectively, systems ranging from photonic crystal and phononic crystals to mean-field electrons in the tight-binding approximation.
    }
\end{figure}

The goal of this paper is to (i) systematically construct families of Hamiltonians endowed with a given duality and (ii) provide a universal description of Hamiltonians families near self-dual points.
We are particularly interested in the consequences of dualities on the properties of spatially extended \notemph{materials} (and metamaterials). 
To handle such systems, we focus on tight-binding models (also known as coupled-mode theories), which describe a collection of coupled harmonic oscillators ranging from atomic orbitals, mechanical resonators to optical cavities, see Fig.~\ref{figure_resonators_dualities}. 
These models provide a somewhat universal effective description of linear systems, provided that one doesn't insist that each individual oscillator should necessarily correspond to an identifiable physical entity. Most crucially from an engineering perspective, systematic methods have been developed to solve the inverse problem of designing realizable systems that correspond to a given tight-binding Hamiltonian~\cite{Matlack2018,Fruchart2018b,Ozawa2019,Cooper2019,Bloch2008}.

\section{Dualities in continuous families of Hamiltonians}
\label{dualities_continuous_families}

Our topic requires some conscious attention to the parameters $p$, and we denote by $P$ the parameter space in which they live. We call $H(p)$ the Hamiltonian of the system with the parameters $p$, and ask that the function $p \mapsto H(p)$ is smooth.

We say that the family of Hamiltonians $H(p)$ is endowed with a duality when~\footnote{In this work, we focus on dualities as defined by Eq.~\eqref{duality_def}.
This definition is inspired by similar (but not identical) kinds of dualities crucial in statistical physics~\cite{Savit1980}, mechanics~\cite{Crapo1993,Zhou2019}, condensed matter physics~\cite{Senthil2004,Zaanen2015,Senthil2019} or high-energy physics~\cite{Hull1995,Maldacena1999}.}
\begin{equation}
    \label{duality_def}
    \hat{U} \, \hat{H}(f(p)) \, \hat{U}^{-1} = \hat{H}(p)
\end{equation}
where $\hat{U}$ is a unitary operator and $f$ is a map from the parameter space $P$ to itself~\footnote{We assume that $f$ is a homeomorphism: a continuous bijection with a continuous inverse.}.
We emphasize that the duality is determined not only by $\hat{U}$ but also by~$f$.
The fixed points of the map $f$ are called self-dual points. 
At a self-dual point, $\hat{H}(p) \equiv \hat{H}(f(p))$ and the duality reduces to a symmetry.

As a whole, the duality acts on the composite space $\mathcal{S} = \text{End}(\mathcal{H}) \times P$, where $\mathcal{H}$ is the Hilbert space of physical states and $\text{End}(\mathcal{H})$ is the space of linear maps on $\mathcal{H}$ that contains all possible Hamiltonians~\footnote{
    We have made the simplifying assumption that the space of physical states $\mathcal{H}$ doesn't depend on the parameters. 
    This is not necessarily true: in general, $P \times \mathcal{H}$ may be replaced by vector bundle $\mathcal{E}$ over the parameter space~$P$, and~$\mathcal{S}$ may be replaced by the endomorphism bundle~$\text{End}(\mathcal{E})$. 
    In this case, dualities can be expressed in terms of equivariant vector bundles (see Refs~\cite{Segal1968,Merkurjev2005} and references therein for definitions).
Starting with an abstract duality group $G$, we ask that $\mathcal{E}$ is a $G$-equivariant vector bundle.
    The endomorphism bundle is then also $G$-equivariant with the corresponding adjoint action.
    Let us give a bit more detail. 
    The group elements $g \in G$ are an abstract version of the different dualities acting on our system; we still have to specify how they act.
    To do so, we consider an action $f : G \times P \to P$ of $G$ on the parameter space.
    Each $g \in G$ gives rise to a function $p \to f_g(p)$ on $P$. 
    This specifies how parameters change under the duality~$g$.
    Then, we define a linear action $U : G \times \mathcal{E} \to \mathcal{E}$ of $G$ on the vector bundle $\mathcal{E}$ of physical states. 
    It specifies how states are transformed.
    We impose that $U_g = U(g,\cdot)$ is a linear map from the fiber $\mathcal{E}_p$ over $p$ to the fiber $\mathcal{E}_{f_g(p)}$ over~$f_g(p)$.
    This is the equivariance condition, which simply means that a state $\psi$ of the system with parameters~$p$ is mapped to a state $U_g \psi$ of the system with parameters~$f_g(p)$.
    We can deduce how operators transform from the way states transform. 
    Accordingly, an operator $H$ acting on the states of the system with parameters $p$ is mapped to $U_g H U_g^{-1}$, which is an operator acting on the states the system with parameters $f_g(p)$.
}.
The action on this composite space is generated by $\mathcal{U} = (f, \text{ad}_U)$ 
(the function $\text{ad}_U$ is defined by $\text{ad}_U(H) = U H U^{-1}$).
The order of an operation $\mathcal{O}$ is the smallest integer $n$ such that $\mathcal{O}^n = \Id$ is the identity (the order is set to $\infty$ when no such integer exists).
As a duality $\mathcal{U}$ is composed of two pieces $f$ and $U$, we can define two separate orders.
First, let $m$ be the order of the duality map, i.e. the smallest integer such that $f^{\circ m}(p) = p$ for all $p$ (here, $f^{\circ m} = f \circ f \circ \dots \circ f$ is the $m$th iterate of the function $f$, where $\circ$ represents composition).
Second, let $n$ be the order of the duality operator, such that $\text{ad}_U^{\circ n} = \text{id}$~\footnote{We can define a separate order $n'$ such that $U^{n'} = \Id$ is the identity operator. In general, $n \neq n'$.}.
In the following, we assume that $m = n$, and refer to this integer as \emph{the} order of the duality.
When $m \neq n$, the Hamiltonian is either endowed with additional symmetries, or there are redundancies in the parameters (i.e., there are equal Hamiltonians for different values of parameters). 
For simplicity, we do not consider these situations in this work. Besides, we assume that the order $n=m$ of the duality is finite~\footnote{As a counter-example, consider the function $f(p) = p^2$ on the parameter space $P = [-1,1]$, which has infinite order. Similarly, consider the diagonal matrix $U = \text{diag}(\ee^{\ii \alpha}, \ee^{\ii \beta})$. Its action by conjugation on a matrix $H$ multiplies the off-diagonal elements of $H$ by $\ee^{\pm \ii (\alpha - \beta)}$ and has infinite order when $\alpha - \beta$ is irrational.}. 

In principle, multiple duality operations can be present.
Consider two pairs $\mathcal{U}_1 = (f_1, \text{ad}_{U_1})$ and $\mathcal{U}_2 = (f_2, \text{ad}_{U_2})$ satisfying Eq.~\eqref{duality_def}. Then, $\mathcal{U}_1 \mathcal{U}_2 = (f_1 \circ f_2, \text{ad}_{U_1} \circ \text{ad}_{U_2})$ must also satisfy Eq.~\eqref{duality_def}. 
Hence, we can see the duality operations $\mathcal{U}_a$ as the representations of an abstract duality group $G$ acting on $\mathcal{S}$.
In this paper, we will focus on the situation in which a single duality operation with finite order $n$ is present; then, $G$ is the cyclic group $C_n$.

\section{Generating dual families on demand}

\subsection{General strategy}
\label{generation_general_strategy}

\def\Gammares{\ensuremath{\Gamma_{0}}}

Given a duality operator $\hat{U}$ (acting on a certain space that is also given), we wish to systematically find the families of Hamiltonians $p \to \hat{H}(p)$ satisfying
\begin{equation}
    \label{duality_equations_commutation}
    \hat{U} \hat{H}(f(p)) = \hat{H}(p) \hat{U}.
\end{equation}
Ideally, we would like an algorithm whose input is composed of a suitable description of the space of states (e.g., a crystal $\mathcal{C}$ with its Bravais lattice and unit cell and the internal degrees of freedom attached to each site) and of the duality operator $\hat{U}$.
Its output should consist in some kind of basis whose elements can be combined to express all the families of Hamiltonians satisfying Eq.~\eqref{duality_equations_commutation}.
Methods and tools performing this task have been developed for spatial symmetries.
This is the main idea of the theory of invariants of solid-state physics~\cite{BirPikus1975,Winkler2003,Willatzen2009,Pikus1961,Luttinger1956}, which builds upon the abstract theory of invariants in abstract algebra~\cite{Weyl1939}.
This approach originated in the study of semiconductors, but it applies to any linear systems, including photonic and phononic crystals~\cite{Sakoda2004,Fruchart2018}.
Besides, several works have recently been devoted to expanding and automating this procedure using computer algebra systems~\cite{Varjas2018,Gresch2018,Chertkov2020}.
Here, we wish to extend this method to dualities.

We will only consider functions $f$ that can smoothly be deformed into isometries by a change of coordinate~\footnote{Namely, we ask that there is another homeomorphisms $g$ such that $g \circ f \circ g^{-1}$ is an isometry, see Ref.~\cite{Kuznetsov2004}.}.
Further, we assume that the duality map $f$ acts as a \emph{linear isometry} on parameter space~\footnote{
Because we have assumed the duality to have finite order, we consider only periodic homeomorphisms (homeomorphisms $f$ such that $f^{\circ m} = \text{id}$ for a finite integer $m$).
(The case of infinite-order dualities is outside of the scope of this work.)
This hypothesis, along with that $f$ is an isometry, gives relatively strong constraints on the duality map.
If we are only interested about local (i.e., not global) properties in parameter space, this implies that $f$ is linear. This can be seen from a series expansion near a self-dual point $p_0$ as follows.
Write $f(p) = f(p_0 + \delta p) = p_0 + F \, \delta p + \mathcal{O}(\delta p^2)$ where $\delta p = p - p_0$. The matrix $F$ satisfies $F^m = \Id$ (because $f^{\circ m} = \text{id}$), and is therefore orthogonal.
The case of global properties is considerably more complicated; however, several results exist in simple cases.
For instance, all the periodic homeomorphism of the real line $\RR$ or on the closed interval $[-1, 1]$ are either the identity map $\text{id}$ or topologically conjugate to the reflection map $x \mapsto -x$. We direct the reader to Ref.~\cite{Constantin2003} and references therein for more details.
}.
Namely, we assume that $f(p) = F p$ in which $F$ is an orthogonal (or unitary) matrix.
This assumption is not as restrictive as it may seem: it turns out that in simple cases, it always holds for dualities of finite order~\footnote{This is a consequence of the Mazur-Ulam theorem and its generalizations, when applicable, as we have assumed that the duality map is an isometry. We direct the reader to Refs.~\cite{Mazur1932,Vaisala2003,Molnar2015} and references therein for more details.}.
It will allow us to use group-theoretical methods to describe the general structure of dualities. 
These will be practically advantageous to obtain explicit parameterized forms of Hamiltonian families with dualities.
The simplest application of this general strategy is presented in Box~\ref{exdualityinterval}, and a concrete example is presented in Box~\ref{exsimpleduality}. The next paragraphs describe the general case.

\begin{example}[label={exdualityinterval}]{dualities in a 1D parameter space}
The simplest situation occurs when there is a single self-dual point. 
This situation can be reduced to the parameter space $P = [-1,1]$ in which the self-dual point is at $p = 0$ (in red), with $f(p) = -p$.
\begin{center}
\includegraphics{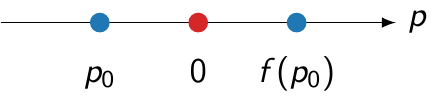}
\end{center}
It turns out that the only other possibility is that the entire interval is self-dual (with $f(p) = p$). 
Consider on the one hand arbitrary odd and even functions $a_{\pm}$ on parameter space, satisfying
\begin{equation}
    \label{example_odd_even_gen}
    a_{\pm}(- p) = \pm a_{\pm}(p)
\end{equation}
and on the other hand, self-dual and anti-self-dual matrices $\hat{H}_{\pm}$ satisfying
\begin{equation}
    \label{example_sd_asd_gen}
    \hat{U} \, \hat{H}_{\pm} \hat{U}^{-1} = \pm \hat{H}_{\pm}.
\end{equation}
Then, the combinations
\begin{equation}
    a_{+}(p) \hat{H}_{+} 
    \quad
    \text{and}
    \quad
    a_{-}(p) \hat{H}_{-} 
\end{equation}
satisfy the duality relation!
We can sum any of such combinations, and the most general family of Hamiltonians with duality then reads
\begin{equation}
    \hat{H}(p) = \sum_{i} a_{+}^{i}(p) \hat{H}_{+}^{i} + \sum_{i} a_{-}^{i}(p) \hat{H}_{-}^{i}
\end{equation}
in which $\hat{H}_{\pm}^{i}$ are linearly independent solutions of Eq.~\eqref{example_sd_asd_gen}, 
and $a_{\pm}^{i}(p)$ are arbitrary functions satisfying Eq.~\eqref{example_odd_even_gen}.
\end{example}

Recall from Sec.~\ref{dualities_continuous_families} that we have introduced an abstract duality group $G$. 
Consider a basis of operators $\hat{H}_{\gamma}^{\mu}$ that transform under an irreducible representation (irrep) $\Gamma_\gamma$ of $G$ (here, $\mu$ labels the different basis matrices of the fixed irrep $\Gamma_\gamma$).
Hence,
\begin{equation}
    \hat{U} \hat{H}_{\gamma}^{\mu} \hat{U}^{-1} = \rho(\gamma)_{\mu \mu'} \hat{H}_{\gamma}^{\mu'}
\end{equation}
in which the matrix $\rho(\gamma)$ depends only on the irreducible representation $\Gamma_\gamma$, and there is an implicit sum over the repeated indices $\mu'$.
Similarly, consider a basis of functions on parameter space $a_{\gamma}^\mu : P \to P$ that transform under the conjugate irrep $\Gamma_\gamma^*$.
This means that
\begin{equation}
    a_{\gamma}^{\mu}(f(p)) = \overline{\rho(\gamma)}_{\mu \mu'} \; a_{\gamma}^{\mu'}(p)
\end{equation}
in which the overline corresponds to complex conjugation.
(Note that under our assumptions, the duality map $f$ is a linear map $f(p) = F p$ in which $F$ is a matrix acting on parameters.)
We can now combine both ingredients into
\begin{equation}
    \hat{H}_{\gamma}(p) = a_{\gamma}^{\mu}(f(p)) \hat{H}_{\gamma}^{\mu}.
\end{equation}
We then find
\begin{equation}
   \hat{U} \hat{H}_{\gamma}(F p) \hat{U}^{-1} = \hat{H}_{\gamma}(p)
\end{equation}
using that $\rho(\gamma)$ is unitary~\footnote{
We get $\hat{U} \hat{H}_{\gamma}(F p) \hat{U}^{-1} = \delta_{\mu \nu} \, \overline{\rho(\gamma)}_{\mu \mu'} \, \rho(\gamma)_{\nu \nu'} \, a_{\gamma}^{\mu'}(p) \, \hat{H}_{\gamma}^{\nu'}$
and as $\rho(\gamma)$ is unitary, we have $\delta_{\mu \nu} \overline{\rho(\gamma)_{\mu \mu'}} \rho(\gamma)_{\nu \nu'} = [\rho(\gamma)^\dagger \rho(\gamma)]_{\mu' \nu'} = \delta_{\mu' \nu'}$ which gives the result.
}.

Hence, we have constructed a family of functions $\hat{H}(p)$ that satisfies the duality equation \eqref{duality_def} with $f(p) = F p$.
Of course, one could have multiple sets of basis functions (say, polynomials of different orders), and of basis operators.
We have to sum all the combinations with arbitrary coefficients (but the coefficients have, of course, to be the same inside each individual combination).

We can now combine the $\hat{H}_{\gamma}$ corresponding to all the irreps $\Gamma_\gamma$ of $G$ into
\begin{equation}
    \hat{H}(p) = \sum_{\gamma} a_\gamma \hat{H}_{\gamma}(p)
\end{equation}
which is our generic family of Hamiltonians with duality.

\subsection{Cyclic duality group}

\begin{figure}
    \includegraphics[width=6cm]{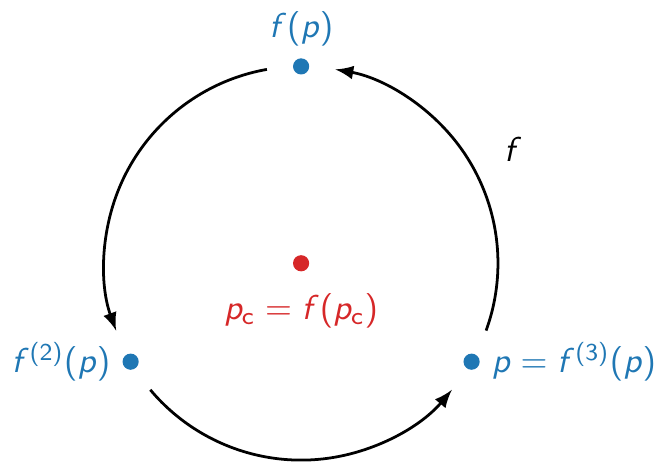}
    \caption{\label{duality_order_three}
    \textbf{Action of an order-three duality on parameter space.}
    Here, we consider a duality action $f$ of order three acting on the 2D plane. 
    We have $f^{(2)} \neq \text{id}$ and $f^{(3)} = \text{id}$.
    In this situation, there is a single self-dual point $p_{\text{c}}$ (in red).
    A typical point $p$ (in blue) has an orbit consisting of three points $p$, $f(p)$, and $f^{(2)}(p)$.
    Applying the duality action a third time maps back to the original point $f^{(3)}(p) = p$.
    }
\end{figure}

Up to here, our discussion applies to any finite group $G$ (and with a few modifications, to any compact group).
We now focus on the situation of a single duality operator with finite order $n \in \mathbb{N}$. 
In this case, $G$ is the cyclic group $C_n$ (Fig.~\ref{duality_order_three} shows an example of order $n=3$ duality).  
The irreps are one-dimensional and the $\rho(\gamma)$ are complex numbers of modulus one.
The irreducible representations of $C_n$ are one-dimensional, and have characters $Z_n^m \equiv \ee^{\ii m 2\pi/n}$.
Accordingly, the basis matrices $\hat{H}_{m}$ for each irreducible representation are characterized by
\begin{equation}
    \label{example_cyclic_ham}
    \hat{U} \, \hat{H}_{m} \hat{U}^{-1} = \ee^{\ii m 2\pi/n} \hat{H}_{m}.
\end{equation}
(Note that this equation could be rewritten without complex numbers, by pairing elements with conjugate characters.)
Following the strategy delinated above, we wish to cancel the phase factor $\ee^{\ii m 2\pi/n}$ by changing the value of the parameters.
To do so, consider the two-component parameter $p=(p_x, p_y)$ that we represent as the complex number $z = p_x + \ii p_y$.
Let us find functions $z \mapsto a_m(z)$ satisfying
\begin{equation}
    \label{example_cyclic_fun}
    a_{m}(\ee^{-\ii m 2\pi/n} z) = \ee^{-\ii m 2\pi/n} a_{m}(z).
\end{equation}
Then, the combinations
\begin{equation}
    \hat{H}(z) = \sum_m a_{m}(z) \hat{H}_{m}
\end{equation}
satisfy the duality relation
\begin{equation}
    \hat{U} \hat{H}( \ee^{-\ii m 2\pi/n} z ) \hat{U} = \hat{H}(z).
\end{equation}

\begin{example}[label={exsimpleduality}]{self-dual and anti-self-dual Hamiltonians}
Consider the unitary matrix
\begin{equation}
    U = \begin{pmatrix} 0 & 1 \\ -1 & 0 \end{pmatrix}.
\end{equation}
We wish to obtain the families of Hamiltonians $H(p)$ satisfying $U H(-p) U^{-1} = H(p)$.
To do so, we first seek a basis for the linear spaces of solutions of the (anti-)self-duality equations $U H U^{-1} = \pm H$.
We find two self-dual basis matrices
\begin{equation}
    \label{exdualityinterval_plus_generators}
    H_{+}^{1} = \begin{pmatrix} 1 & 0 \\ 0 & 1 \end{pmatrix}
    \quad
    H_{+}^{2} = \begin{pmatrix} 0 & -\ii \\ \ii & 0 \end{pmatrix}
\end{equation}
and two anti-self-dual ones,
\begin{equation}
    \label{exdualityinterval_minus_generators}
    H_{-}^{1} = \begin{pmatrix} 1 & 0 \\ 0 & -1 \end{pmatrix}
    \quad
    H_{-}^{2} = \begin{pmatrix} 0 & 1 \\ 1 & 0 \end{pmatrix}.
\end{equation}

Then, we look for functions satisfying Eq.~\eqref{example_odd_even_gen}. 
Let us use polynomials: $a_{\pm}$ are respectively polynomials with only even or odd powers of $p$.

Finally, we combine everything together to write
\begin{equation}
\label{exdualityinterval_hamiltonian_family}
\begin{split}
    H(p) = 
      ([a_{+}^{1}]_0 + [a_{+}^{1}]_2 \, p^2 + \cdots) H_{+}^{1} \\
    + ([a_{+}^{2}]_0 + [a_{+}^{2}]_2 \, p^2 + \cdots) H_{+}^{2} \\
    + ([a_{-}^{1}]_1 \, p + [a_{-}^{1}]_3 \, p^3 + \cdots) H_{-}^{1} \\
    + ([a_{-}^{2}]_1 \, p +[a_{-}^{2}]_3 \, p^3 + \cdots) H_{-}^{2}
\end{split}
\end{equation}
in which all the $[a_{\pm}^{a}]_i$ are independent (arbitrary) complex numbers.
\end{example}

\subsection{Spatially periodic metamaterials}

We now focus on spatially periodic systems (crystals) which are obtained by repeating a unit cell over a Bravais lattice $\Gamma$ (see Appendix~\ref{app_bloch_representations} for details).
To do so, we consider Hamiltonians of the form
\begin{equation}
	\label{generic_hamiltonian}
    \hat{H}(p) = \sum_{\gamma \in \Gammares} h(\gamma, p) \hat{T}(\gamma)
\end{equation}
in which $\hat{T}(\gamma)$ is translation operator by a Bravais lattice vector $\gamma$, $h(\gamma)$ is a finite-dimensional matrix acting on the degrees of freedom in the unit cell~\footnote{To make contact with equivalent notations, note that the translation operator can be written $\hat{T}(\gamma) = \sum_{x \in \mathcal{C}} \ket{x+\gamma}\!\bra{x}$ or equivalently $\hat{T}(\gamma) = \sum_{x \in \mathcal{C}} \hat{c}^\dagger_{x+\gamma}\, \hat{c}_x$ where $\ket{x}$ is a state fully localized at the point $x$ of the crystal, and $\hat{c}_x^{(\dagger)}$ the corresponding annihilation (creation) operator.
Given a basis $\ket{e_i}$ of the Hilbert space of the degrees of freedom in the unit cell, we can further decompose $h(\gamma, p) = \sum_{} h_{i j}(\gamma, p) \ket{e_i}\!\bra{e_j}$ in which the matrix elements $h_{i j}(\gamma, p)$ are now scalars.
The Hamiltonians $\hat{H}(p)$ act on the Hilbert space spanned by states of the form $\ket{x} \otimes \ket{e_i}$.
}.
All translation invariant Hamiltonians can be written in this form.
For convenience, we also assume that the connections between the oscillators have a finite range (this hypothesis is not crucial, but simplifies the manipulations). 
Correspondingly, the sum is restricted to a finite subset $\Gammares \subset \Gamma$ of the Bravais lattice $\Gamma$, centered at the origin.

We also decompose the duality operator as
\begin{equation}
    \hat{U} = \sum_{\gamma,\gamma' \in \Gamma} u(\gamma, \gamma') \ket{\gamma} \bra{\gamma'}
\end{equation}
in which $u(\gamma, \gamma')$ are again finite-dimensional matrices, and in which $\ket{\gamma}$ represents the function in $\ell^2(\Gamma)$ having value one at $\gamma$ and zero elsewhere~\footnote{We can also write more explicitly
\begin{equation}
    \hat{U} = \sum_{{\protect\substack{\gamma,\gamma' \in \Gamma \\ \xi,\xi' \in \Fun}}} u_{\xi',\xi}(\gamma', \gamma) \ket{\gamma', \xi'} \bra{\gamma, \xi}
\end{equation}
in which $\ket{\gamma, \xi}$ are a basis of $\ell^2(\Cr)$ of functions $x \mapsto \delta(x - (\gamma + \xi))$ (for $\gamma \in \Gamma$ and $\xi \in \Fun$, where $\Fun$ is a fundamental domain) having value one at $\gamma + \xi \in \Cr$ and zero elsewhere.
}.

Equation~\eqref{duality_equations_commutation} can be decomposed over the translation operators and transformed to a finite number of linear relations between the matrix elements $h_{i j}(\gamma, p)$ and the dual matrix elements $h_{i j}(\gamma, f(p))$, gathered in a linear equation of the form
\begin{equation}
    \label{linear_eq_duality_AB}
    \mathcal{A}[\hat{U}] \, \vec{h}(p) = \mathcal{B}[\hat{U}] \, \vec{h}(f(p))
\end{equation}
in which the matrix elements $h_{i j}(\gamma, p)$ are all gathered in the vector $\vec{h}(p)$, etc. and 
in which $\mathcal{A}[\hat{U}]$ and $\mathcal{B}[\hat{U}]$ are known matrices (that depend on the duality operator)~\footnote{
One can further convert Eq.~\eqref{linear_eq_duality_AB} into the homogeneous linear system $[\mathcal{A}, \mathcal{B}] \, [\vec{h}(p), \vec{h}(f(p))]^T = 0$.
The vectors and matrices are finite, provided that we restrict the model to have finite range connections (corresponding to the restriction to $\Gammares$).
}.
Equation~\eqref{linear_eq_duality_AB} can be solved (numerically or symbolically, see e.g. Refs.~\cite{Geddes1992,Press2007}), giving $\vec{h}(f(p))$ as a function of $\vec{h}(p)$ (or conversely).

\subsection{Obtaining explicit parameterized families}

The general strategy developed up to here doesn't make any assumptions on the duality, but it leaves us with an set of equations relating $\vec{h}(f(p))$ and $\vec{h}(p)$.
In practice, it is convenient to obtain explicit parameterized forms of the Hamiltonian families. 
To do so, we will restrict our attention to dualities in which the duality map $f$ is linear (as well as all dualities that can be obtained from those by reparameterization).
This hypothesis allows us to use the structure described in Section.~\ref{generation_general_strategy} to obtain an systematic way of explicitly constructing the families of Hamiltonians with a duality~\footnote{It is still possible to directly solve the duality equation when $f$ is non-linear, but we don't have a systematic way of writing an explicit parametrization.}. 

To do so, we will first make change of variables to make parameter space as symmetric as possible (we assume that the duality map $f$ becomes an isometry); second, we need to use a basis of functions adapted to the symmetry. 
We then specialize the strategy of Section.~\ref{generation_general_strategy} to Eq.~\eqref{linear_eq_duality_AB}.
To get an explicit form, one might further parameterize these functions, e.g. through a series expansion, a Fourier or Chebyshev decomposition, etc. (see an example with polynomials in Boxes~\ref{exsimpleduality} and \ref{exdualityinterval}).

\begin{example}{explicit parameterization for dualities on the line.}
We take as a parameter space the real line $P=\mathbb{R}$, and $f(p) = - p$. 
There is a single self-dual point at $p=0$.
Consider on the one hand arbitrary odd and even functions $a_{\pm}$ on parameter space, satisfying
\begin{equation}
    \label{example_odd_even}
    a_{\pm}(f(p)) = \pm a_{\pm}(p)
\end{equation}
and on the other hand, self-dual and anti-self-dual vectors $\vec{h}_{\pm}$, satisfying
\begin{equation}
    \label{example_sd_asd}
    \mathcal{A}[\hat{U}] \, \vec{h}_{\pm} = \pm \mathcal{B}[\hat{U}] \, \vec{h}.
\end{equation}
(Note that $\vec{h}_{\pm}$ do not depend on the parameters.)
Then, the combinations
\begin{equation}
    a_{+}(p) \vec{h}_{+} 
    \quad
    \text{and}
    \quad
    a_{-}(p) \vec{h}_{-} 
\end{equation}
satisfy the duality relation.
The most general family of Hamiltonians with duality then reads
\begin{equation}
    \vec{h}(p) = \sum_{i} a_{+}^{i}(p) \vec{h}_{+}^{i} + \sum_{i} a_{-}^{i}(p) \vec{h}_{-}^{i}.
\end{equation}
In this equation, $\vec{h}_{\pm}^{i}$ form a basis of independent solutions of the linear Eq.~\eqref{example_sd_asd}, 
and $a_{\pm}^{i}(p)$ are arbitrary functions satisfying Eq.~\eqref{example_odd_even} (i.e., odd and even functions).
We can then parameterize the functions of $a_{\pm}^{i}(p)$, for instance by performing a series expansion near the self-dual point, leading to
\begin{equation}
    a_{+}^{i}(p) = \sum_n \alpha_n^{i} p^{2 n}
\end{equation}
and 
\begin{equation}
    a_{-}^{i}(p) = \sum_n \beta_n^{i} p^{2 n+1}.
\end{equation}
(Here, $\alpha_n^{i}$ is a coefficient but $p^{2 n}$ is the $2n$th power of $p$.)
Let us summarize: we first have to solve Eq.~\eqref{example_sd_asd} to obtain the basis elements $\vec{h}_{\pm}^{i}$.
Then, we can choose arbitrary coefficients $\alpha_n^{i}$ and $\beta_n^{i}$ to get a family of Hamiltonians depending on the control parameter $p$ that satisfies the duality equation.
In Sec.~\ref{handcrafted}, we illustrate this strategy on a concrete example of duality.
\end{example}

\subsection{Bloch representation}

In spatially periodic systems, it is convenient to use the Bloch (momentum space) representation. We refer to Appendix~\ref{app_bloch_representations} for details. In short, each Hamiltonian operator $\hat{H}(p)$ (for a fixed $p$) is mapped to a continuous family of matrices $k \mapsto H(p, k)$ with the quasi-momentum $k$ in the Brillouin zone (a torus $T^d$).
We emphasize that the quasi-momentum $k$ does not play the same role as the external parameter $p$: the entire family $k \mapsto H(k)$ for $k$ in the Brillouin zone describes a single physical system, while different values of $p$ correspond to different physical systems.
In general, the duality changes the quasi-momentum $k$ to $\mathcal{O} k$, in which $\mathcal{O}$ is an orthogonal matrix.
We can then write the equivalent of Eq.~\eqref{duality_def} for the Bloch Hamiltonians~\footnote{We have shortened to $U(\mathcal{O} k)$ the momentum-dependent duality operator $U(k, \mathcal{O} k)$. This is a choice: we could have called the same quantity $U(k)$.}
\begin{equation}
    \label{duality_bloch}
    U(\mathcal{O} k) \, H(f(p), \mathcal{O} k) U^{-1}(\mathcal{O} k) = H(p, k).
\end{equation}

\subsection{Example: duality in a 1D crystal}
\label{handcrafted}

Consider first a 1D crystal with two degrees of freedom per unit cell, and the duality defined in Bloch space by $f(p) = -p$, $\mathcal{O} k = - k$ in Eq.~\eqref{duality_bloch}, as well as
\begin{equation}
	\label{handcrafted_example_U}
    U(k) = \ii \sigma_y = \left(\begin{matrix} 0 &  1 \\ - 1 & 0 \end{matrix} \right).
\end{equation}
(Here, $k \in S^1$ is the one-dimensional quasi-momentum.)

In real space, this corresponds to the operator
\begin{equation}
    \hat{U} = \ii \sigma_y \sum_{\gamma} \ket{T(-\gamma)} \bra{T(\gamma)}
\end{equation}
The second part is due to the action $\mathcal{O} k = - k$ on momentum space.

We now consider Bloch Hamiltonians of the form
\begin{equation}
	\label{handcrafted_bloch_hamiltonian}
    H(p, k) = H^{(0)}(p) + H^{(1)}(p) \ee^{-\ii k} + H^{(-1)}(p) \ee^{\ii k}.
\end{equation}
The duality condition Eq.~\eqref{duality_bloch} applied to Eq.~\eqref{handcrafted_bloch_hamiltonian} gives
\begin{subequations}
\begin{align}
    \label{handcrafted_duality_H0} U H^{(0)}(-p) U^{-1} &= H^{(0)}(p) \\
    \label{handcrafted_duality_H1pm} U H^{(1)}(-p) U^{-1} &= H^{(-1)}(p)
\end{align}
\end{subequations}
in which $U \equiv U(k)$ defined in Eq.~\eqref{handcrafted_example_U}.

To handle Eq.~\eqref{handcrafted_duality_H0}, let us follow Box~\ref{exdualityinterval} and look for a basis $H^i_{\pm}$ of (anti-)self-dual solutions to $U H^i_{\pm} U^{-1} = \pm H^i_{\pm}$.
This has been done in Box~\ref{exsimpleduality}, and $H^{(0)}(p)$ is given by the RHS of Eq.~\eqref{exdualityinterval_hamiltonian_family}.
To satisfy Eq.~\eqref{handcrafted_duality_H1pm}, we can simply take the matrix elements $H_{i j}^{(1)}(p)$ to be arbitrary smooth functions (without any constraint), and Eq.~\eqref{handcrafted_duality_H1pm} then fully determines $H^{(-1)}(p)$ as a function of $H^{(1)}(-p)$.

\medskip

Let us be more ambitious and add an additional specification: we now want $H(p)$ to be Hermitian.
This requires 
\begin{equation}
H^{(0)} = [H^{(0)}]^\dagger 
\quad
\text{and} 
\quad
H^{(1)}(p) = [H^{(-1)}(p)]^\dagger.
\end{equation}
in addition to Eqs.~\eqref{handcrafted_duality_H0} and \eqref{handcrafted_duality_H1pm}.

Let us first worry about $H^{(0)}$. 
We can first impose the constraint Eq.~\eqref{handcrafted_duality_H0} as explained above. 
As the generators $H_{\pm}^a$ in Eqs.~(\ref{exdualityinterval_plus_generators}-\ref{exdualityinterval_minus_generators}) are Hermitian, the Hermiticity of $H^{(0)}$ is then ensured by taking 
the coefficients $[a_{\pm}^{a}]_i \in \mathbb{R}$ in Eq.~\eqref{exdualityinterval_hamiltonian_family}.
The second condition requires a bit more work: we are not anymore free to choose an arbitrary function $H^{(1)}(p)$.
Instead, we have to find all families of matrices satisfying
\begin{equation}
	U H^{(1)}(-p) U^{-1} = [H^{(1)}(p)]^\dagger
\end{equation}
and then set $H^{(-1)}(p) = [H^{(1)}(p)]^\dagger$.
Proceeding in a similar fashion as in Box~\ref{exsimpleduality}, we look for a basis $\mathcal{B}_{\pm} = (H_{\pm}^j)_j$ of the vector spaces of matrices $H_{\pm}$ satisfying
\begin{equation}
 	U H_{\pm} U^{-1} = \pm [H_\pm(p)]^\dagger
\end{equation} 
(Note that these $H_{\pm}^j$ are not the same as the ones in Box~\ref{exsimpleduality}.)
We find $\mathcal{B}_{+} =(\sigma_0, \ii \sigma_1, \sigma_2, \ii \sigma_3)$ and $\mathcal{B}_{-} = (\ii \sigma_0, \sigma_1, \ii \sigma_2, \sigma_3)$, where $\sigma_j$ are Pauli matrices.
These matrices are then combined with odd/even polynomials (with real coefficients) to construct $H^{(1)}(p)$.

\begin{figure}
    \includegraphics{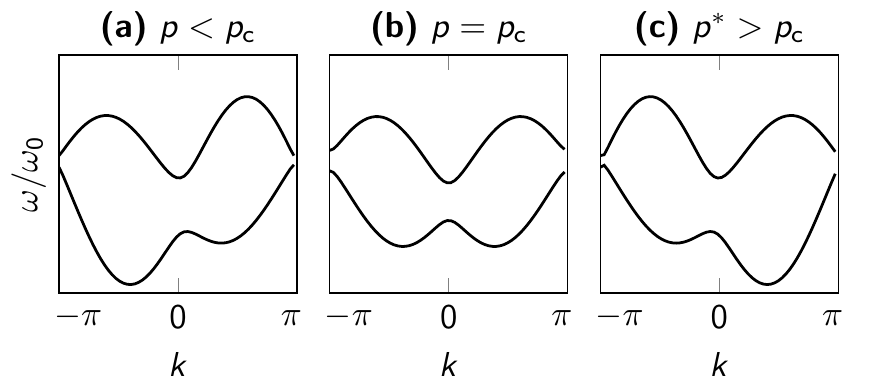}
    \caption{\label{figure_toy_example}
    \textbf{A duality with no degeneracy.} We have used a random polynomial of order $2$ (random coefficients $[a_{\pm}^{a}]_i$ in Eq.~\eqref{exdualityinterval_hamiltonian_family} and its equivalents).
    In this example, there ain't no bosonic TRI for to give you no degeneracy.
    }
\end{figure}

Putting everything together, we obtain Hermitian Hamiltonians Eq.~\eqref{handcrafted_bloch_hamiltonian} satisfying the duality condition Eq.~\eqref{duality_bloch} with the duality operator defined by Eq.~\eqref{handcrafted_example_U}.
In Fig.~\ref{figure_toy_example}, we show the band structure of one of these Hamiltonians, chosen at random.
In this figure, we note that (i) the band structures of dual systems (a) and (c) are obtained by applying the transformation $k \to -k$ and (ii) the self-dual band structure is symmetric under $k \to -k$.
As we shall see in the next paragraph, richer consequences arise when the duality is combined with another constraints.

\begin{figure}
    \includegraphics{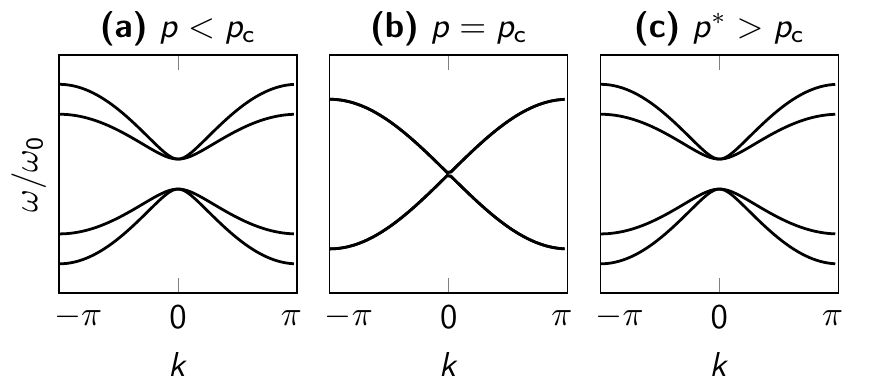}
    \caption{\label{figure_example_1D_bs}
    \textbf{A one-dimensional example of duality.} A $4 \times 4$ example in 1D.
    The systems (a) and (c) are dual to each other, while the system (b) is self-dual.
    The Hamiltonians $H(p)$ in the family are endowed with the duality in Eq.~\eqref{duality_operator_for_examples} and with bosonic TRI (the physical space Hamiltonians are real-valued).
    The coefficients of the Hamiltonian are choosen in order to have a Dirac cone at $k = 0$. 
    Because of the combination of the self-duality with bosonic TRI, the spectrum (b) is two-fold degenerate.
    }
\end{figure}

\begin{figure}
    \includegraphics{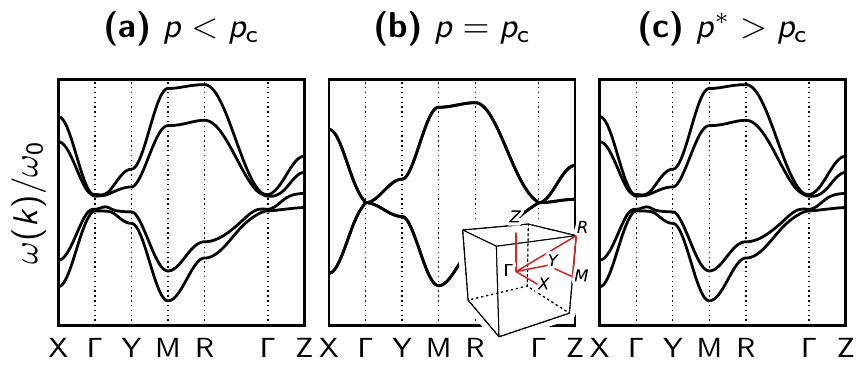}
    \caption{\label{figure_band_structures_3D}
    \textbf{A three-dimensional example of duality.} A $4 \times 4$ example in 3D.
    The systems (a) and (c) are dual to each other, while the system (b) is self-dual.
    We consider a 3D cubic lattice with four degrees of freedom per unit cell. 
    Each unit cell is coupled to the six nearest unit cells with an initially arbitrary $4 \times 4$ hopping matrix.
    The duality constraint is then applied. 
    In addition, the coefficients of the Hamiltonian are choosen in order to have a degeneracy at $k = 0$. 
    The Hamiltonians $H(p)$ in the family are endowed with the duality in Eq.~\eqref{duality_operator_for_examples} and with bosonic TRI (the physical space Hamiltonians are real-valued).
    Because of the combination of the self-duality with bosonic TRI, the spectrum (b) is two-fold degenerate.
    }
\end{figure}

\subsection{Other Short Stories}

We now show the results of the procedure described in section~\ref{generation_general_strategy}.
Before, we must make a few choices: what duality operator $\hat{U}$, what parameter space $P$, and what duality map $f$ shall we impose?
We are free to choose~\footnote{This is contingent on the existence of free will. We refer to the articles~\cite{Conway2006,Chiang2005,Aaronson2016} for discussions.} any unitary $\hat{U}$ and any homeomorphism $f$.
The most simple dualities are associated with a one-dimensional parameter space $P = \mathbb{R}$ and $f(p) = -p$.
We will restrict our attention to this case.
We would also like to find duality operator $\hat{U}$ as simple as possible, but we ask the antiunitary operator $\hat{A} = \hat{U} \mathscr{K}$ (obtained by combining the duality operator $\hat{U}$ with complex conjugation $\mathscr{K}$) squares to $\hat{A}^2 = - 1$.
In this paragraph, we will constrain all tight-binding Hamiltonians to be real-valued (equivalently, when they commute with $\mathscr{K}$).
These requirements guarantee (through Kramers theorem) that the spectra of self-dual Hamiltonians are two-fold degenerate everywhere in the Brillouin zone.

\medskip

This leads us to consider the duality transformation
\begin{equation}
	\label{duality_operator_for_examples}
    \hat{U} = \begin{pmatrix} \ii \sigma_y & 0 \\ 0 & \ii \sigma_y \, T(a_1) \end{pmatrix} \;
    \hat{\mathcal{I}} 
\end{equation}
in which
\begin{equation}
	\label{bravais_lattice_inversion}
	\hat{\mathcal{I}} = \sum_{\gamma} \ket{-\gamma} \bra{\gamma}.
\end{equation}
In momentum space, this corresponds to Eq.~\eqref{duality_bloch} with
\begin{equation}
    U(k) = \begin{pmatrix} \ii \sigma_y & 0 \\ 0 & \ii \sigma_y \, \ee^{-\ii k \cdot {a}_1} \end{pmatrix}
    \quad
    \text{and}
    \quad
    g(k) = - k.
\end{equation}

As we have seen in the previous example (Sec.~\ref{handcrafted} and Fig.~\ref{figure_toy_example}), it is possible to obtain families of Hamiltonians (with dualities) that have no striking feature: the bands are typically disconnected from each other. 
When additional symmetries or constraints are imposed (or by accident), it is also possible to have degeneracies (like Dirac or Weyl points) in a self-dual band structure. 
We illustrate this feature in Fig.~\ref{figure_example_1D_bs} for a 1D system, and in Fig.~\ref{figure_band_structures_3D} for a 3D system.

\section{Dualities in constrained linear systems}
\label{constrained_systems}

In the previous sections, we have assumed that all tight-binding Hamiltonians are available to us, perhaps up to some linear constraints such as asking for Hermitian or real-valued Hamiltonians.
In particular, we have heavily relied on the hypothesis that linear combinations of physically relevant Hamiltonians are still physically relevant Hamiltonians.
Unfortunately, this is not always the case.
As an example, consider the dynamical matrix $\hat{D}$ describing the vibrations of a set of massive particles arranged on a $d$-dimensional crystal and ruled by Newton equations
\begin{equation}
	\partial_t^2 \phi = - D \phi
\end{equation}
in which $\phi$ are the displacements of the masses with respect to their equilibrium positions (see section~\ref{dynamical_matrices} below for details).
Even though the dynamical matrices $D_1$ and $D_2$ might describe two perfectly reasonable networks of particles connected by springs, there is no reason why $D_1 + D_2$ should necessarily describe yet another networks of spring-connected particles~\footnote{It is even dubious that the notation $D_1 + D_2$ has any meaning, because $D_1$ and $D_2$ act on different spaces (that are not canonically isomorphic).}. 
More generally, one might want to consider systems in which only a certain portion of parameter space is accessible, either for fundamental or for practical reasons.
The purpose of this section is to tackle the analysis of dualities in this class of systems.

Deprived from the powerful tools of linear algebra and group theory, we are left with no other choice than to resort to numerical analysis.
To do so, we start by reformulating the duality condition \eqref{duality_def} as an optimization problem.
First, we define
\begin{equation}
	\label{function_F_def}
	\mathcal{F}(p,p') = \hat{U} \, \hat{H}(p') \, \hat{U}^{-1} - \hat{H}(p)
\end{equation}
which is a function defined on the doubled parameter space $P \times P$.
Indeed, Eq.~\eqref{duality_def} is satisfied whenever
\begin{equation}
	\mathcal{F}(p,f(p)) = 0.
\end{equation}
We define $s = (p,p')$ and interpret $\mathcal{F}(s) \equiv \mathcal{F}(p,p')$ as a vector composed of all its matrix elements~\footnote{
In practice, it is convenient to focus on the cases in which there is a finite number of translation operators in Eq.~\eqref{function_F_def} (see the discussion after Eq.~\eqref{generic_hamiltonian}), so that
\begin{equation}
	\mathcal{F}(s) = \sum_{\gamma \in \Gamma} \mathcal{F}(s; \gamma) T(\gamma)
\end{equation}
is a finite sum, in which $\mathcal{F}(s; \gamma)$ are finite-dimensional matrices.}.
Conversely, any $s$ such that $\mathcal{F}(s) = 0$ is a pair of dual parameters. 
We can promote this property as a definition, and (re)define self-dual points $p$ as dual pairs of the form $s=(p,p)$.
This updated definition purposefully leaves out the function $f$ in Eq.~\eqref{duality_def}.

This reformulation affords us two things. 
First, we can look for pairs of dual parameters using optimization or root-finding algorithms (see e.g. Ref.~\cite{Press2007}) by minimizing (or directly finding roots of) the duality potential 
\begin{equation}
	\label{duality_potential}
	V = \lVert \mathcal{F}(s) \rVert^2.	
\end{equation}
Second, we can find neighboring pairs of dual parameters from known ones using continuation (path following) algorithms~\cite{Allgower2003,Kuznetsov2004}.
As a particular case, we can also look for self-dual points and continue them by restricting to diagonal parameters $s=(p,p)$.

\subsection{Minimization: finding dual pairs from scratch}

We simply look for local minima of the duality potential
\begin{equation}
	s^* \in \operatornamewithlimits{Argmin}_s \lVert \mathcal{F}(s) \rVert^2.
\end{equation}
Importantly, we are only interested in the minima $s^*$ that saturate the bound, i.e. those for which $\mathcal{F}(s^*) = 0$. 
It is possible to use standard minimization algorithms and discard the other minima, or to directly use root-finding algorithms (see e.g. Ref.~\cite{Press2007}).
An illustration in a simple mechanical system is presented in Fig.~\ref{figure_self_duality_potential_mechanics_1D} of the next section.

\subsection{Continuation: finding dual pairs from a known one}

\begin{figure}
    \includegraphics[width=6cm]{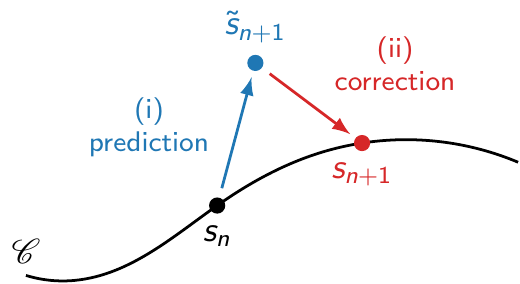}
    \caption{\label{numerical_continuation_fig}
    \textbf{Numerical continuation.}
    Let us assume that the solutions of the equation $F(s) = 0$ form a curve $\mathscr{C}$.
    Starting from a point $s_0$ in this curve (so $F(s_0) = 0$), we want to find the whole curve.
    This is the job of numerical continuation (or numerical path following) algorithms.
    The main idea consists in combining (i) a prediction step (in blue) that starts from a point $s_{n}$ on the curve to suggest a new point $\tilde{s}_{n+1}$ that is a bit off the curve, but in the good direction with (ii) a correction step (in red) that projects $\tilde{s}_{n+1}$ to a new point $p_{n+1}$ actually on the curve.
    }
\end{figure}

When a pair of dual points is known, it is possible to use numerical continuation algorithms~\cite{Allgower2003,Kuznetsov2004} in order to find neighbouring ones. The general strategy is illustrated in Fig.~\ref{numerical_continuation_fig}.
Start from a point $s_n$ such that $\mathcal{F}(s_n) = 0$, we perform a series expansion
\begin{equation}
	\mathcal{F}_I(s) = \mathcal{F}_I(s_n) + F_I^j \, \delta s_j + \mathcal{O}(\delta s^2)
\end{equation}
in which $\delta s = s - s_n$, the index $I$ labels the components of the vector $\mathcal{F}$, the index $j$ labels the components of the doubled parameter $s$, and we have defined
\begin{equation}
	F_I^j = \frac{\partial \mathcal{F}_I}{\partial s_j}
\end{equation}
evaluated at $s_n$.
As $\mathcal{F}_I(s_n) = 0$ by hypothesis, a first-order prediction of the next point of the curve is
\begin{equation}
	\tilde{s}_{n+1} = s_n + \epsilon \, \delta \tilde{s}
\end{equation}
in which $\epsilon$ is a small parameter characterizing the size of the steps, and $\delta \tilde{s}$ is a normalized solution of the linear equations
\begin{equation}
	F_I^j \delta \tilde{s}_j = 0
	\quad
	\text{for all $I$.}
\end{equation}
Note that the space of solutions of this equation may have more than one dimension (this occurs if the manifold of solution is not a curve, but e.g. a surface, or when there are branching points). 
In this case, one has to choose a direction, and different choices will likely lead to different results.
The point $\tilde{s}_{n+1}$ (in blue in Fig.~\ref{numerical_continuation_fig}) is generally not a solution (namely, $\mathcal{F}(\tilde{s}_{n+1}) \neq 0$). 
To get an actual solution (if it exists), we use a correction step that minimizes $\lVert \mathcal{F}(s) \rVert^2$ starting from the initial point $\tilde{s}_{n+1}$ (e.g. with Newton iterations).
In this way, we obtain a new point on the curve (in red in Fig.~\ref{numerical_continuation_fig}) 
\begin{equation}
	s_{n+1} = \operatornamewithlimits{argmin}_{s \leftarrow \tilde{s}_{n+1}} \lVert \mathcal{F}(s) \rVert^2.
\end{equation}
This process can then be iterated.

\subsection{Example: dynamical matrices of mechanical systems}
\label{dynamical_matrices}

\def\eq{\text{eq}}

\subsubsection{General considerations}

Consider a set of massive particles arranged on a $d$-dimensional crystal and interacting via short-range pairwise potentials represented by springs.
The time evolution of the positions $x$ and momenta $\pi$ of the particles is described by the (classical) Hamiltonian
\begin{equation}
\label{hamiltonian_function_mechanical_system}
\mathcal{H}(p,x) = \sum_i \frac{\pi_i^2}{2 m_i} + \frac{1}{2} \sum_{i,j} \frac{k_{i j}}{2} \left[ \lVert x_i - x_j \rVert - \ell_{i j}^\eq \right]^2
\end{equation}
The Hamiltonian $\mathcal{H} = T + V$ is invariant under isometry, but its equilibrium state $(x_i^\eq, \pi_i^\eq = 0)$ defining the crystal of interest spontaneously breaks these symmetries.
(The potential $V$ in $\mathcal{H}$ and Eqs.~(\ref{first_derivative_potential}--\ref{force_constant_matrix_eq}) is the second term in Eq.~\eqref{hamiltonian_function_mechanical_system}. It has no relation with the duality potential defined in Eq.~\eqref{duality_potential}.)

To analyze the linear vibrations of the elastic crystal (phonons), we linearize Hamilton canonical equations (or Newton equations) over the equilibrium state.
To do so, let us compute the first derivatives of the potential
\begin{equation}
	\label{first_derivative_potential}
	\frac{\partial V}{\partial x_\ell^\sigma} = \sum_{m} k_{\ell m} (\lVert x_{\ell m} \rVert - \ell_{\ell m} ) \hat{x}_{\ell m}^\sigma
\end{equation}
in which $x_{ij} = x_i - x_j$ and $\hat{x} = x/\lVert x \rVert$. These vanish at equilibrium (else there would be a net force on some particle and no equilibrium).
The second derivatives are
\begin{equation}
\begin{split}
	\label{second_derivative_potential}
	\!\!\!\!\frac{\partial^2 V}{\partial x_i^\mu \, \partial x_j^\nu} = 
	- k_{i j} \left[ \hat{x}_{i j}^\mu \hat{x}_{i j}^\nu + 
\frac{r_{ij} - \ell_{i j}}{r_{ij}} \left[ \delta^{\mu \nu} - \hat{x}_{i j}^\mu \hat{x}_{i j}^\nu  \right]
	\right]
\end{split}
\end{equation}
for $i \neq j$, and in which $r_{ij} = \lVert x_{i j} \rVert$, and for $i = j$
\begin{equation}
	\label{second_derivative_potential_diagonal}
	\frac{\partial^2 V}{\partial x_i^\mu \, \partial x_i^\nu} = - \sum_{k \neq i} \frac{\partial^2 V}{\partial x_i^\mu \, \partial x_k^\nu}.
\end{equation}

When the springs are at rest in the equilibrium configuration ($r_{ij}^\eq = \ell_{i j}$), the second term in Eq.~\eqref{second_derivative_potential} vanishes and we simply have
\begin{equation}
	\label{force_constant_matrix_eq}
	S_{i j}^{\mu \nu} \equiv \left. \frac{\partial^2 V}{\partial x_i^\mu \, \partial x_j^\nu} \right|_\eq
	 = - k_{i j} \, \hat{x}_{i j}^\mu \hat{x}_{i j}^\nu \quad (i \neq j)
\end{equation}
evaluated at the equilibrium configuration (we omitted the $\eq$ labels in the RHS for readability), with the diagonal part is given by Eq.~\eqref{second_derivative_potential_diagonal}, which ensures that the sum rule
\begin{equation}
	\label{force_constant_matrix_sum_rule}
	\sum_j S^{\mu \nu}_{i j} = 0
\end{equation}
is satisfied, which is a manifestation of the translation invariance of the original Hamiltonian and ensures the presence of the Nambu-Goldstone modes originating from its spontaneous breaking (acoustic phonons).

The displacements $u_{i} = x_{i} - x_{i}^{\eq}$ from each particle's equilibrium position then satisfy the equation of motion
\begin{equation}
	\label{newton_displacements}
	M_i \partial_t^2 u_{i}^{\mu} = - S_{i j}^{\mu \nu} u_{j}^{\nu}.
\end{equation}
It is convenient to define $\phi_i^\mu = \sqrt{M_i} u_i^\mu$ so that Eq.~\eqref{newton_displacements} becomes
\begin{equation}
	\partial_t^2 \phi_i^\mu = - D_{i j}^{\mu \nu} \phi_j^\nu
\end{equation}
where we have defined the dynamical matrix $\hat{D}$ by
\begin{equation}
	D_{i j}^{\mu \nu} = \sqrt{M_i^{-1}} S_{i j}^{\mu \nu} \sqrt{M_j^{-1}}
\end{equation}
in which we have assumed that all the masses are positive.

Upon choosing a fundamental domain, we can decompose each point $x_i^\eq$ into $x_i^\eq = \gamma + \delta_n$ in which $\gamma$ is a Bravais lattice vector and $\delta_n$ an element of the fundamental domain.
Then
\begin{equation}
	\partial_t^2 \phi_{m}^{\mu}(\gamma) = - D_{m m'}^{\mu \mu'}(\gamma,\gamma') \phi_{m'}^{\mu'}(\gamma')
\end{equation}
in which $D(\gamma, \gamma') = D(\gamma - \lambda, \gamma' - \lambda)$
for all Bravais lattice vectors $\lambda$ because of the invariance under Bravais lattice translations.

In momentum space (using convention (2) of Appendix~\ref{app_bloch_representations}), we get the momentum space dynamical matrix
\begin{equation}
	D_{i j}^{\mu \nu}(k) = \sum_{\gamma \in \Gamma} \ee^{-\ii k \cdot (\gamma + \delta_j - \delta_i)}  D_{i j}^{\mu \nu}(0, \gamma)
\end{equation}

Equations~\eqref{force_constant_matrix_eq} and~\eqref{force_constant_matrix_sum_rule} show that not all matrices can be interpreted as the dynamical matrix of a system of particles connected by springs (even when restricting to symmetric matrices). What's more, the masses and spring constant should be positive in normal circumstances, further restricting the space of allowed matrices.

\begin{figure}
    \includegraphics{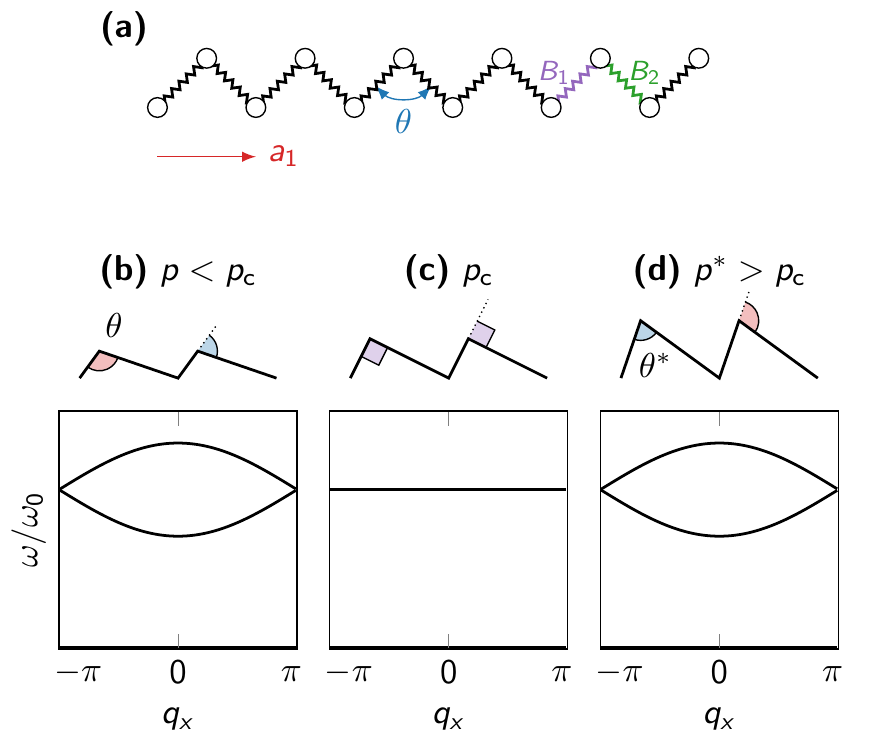}
    \caption{\label{figure_example_mechanics_1D_bs}
    \textbf{Mechanical example in 1D.}
    (a) Family of 1D mechanical systems.
    We have marked the angle $\theta$ between the two inequivalent bonds $B_1$ (in purple) and $B_2$ (in green).
    The primitive vector $a_1$ of the Bravais lattice is drawn in red.
    The positions of the two inequivalent particles are $(x_1,y_1) = (0,0)$ and $(x_2,y_2)$.
    We can choose the parameters to be $p = (\theta, x_2)$. 
    (b-d) Duality and phonon band structures.
    The duality maps $\theta = \pi/2 + \phi$ to $\theta^* = \pi/2 - \phi$.
    We have set $x_2 = \num{0.2}$ and (b) $\theta = \pi/2 - \num{0.3}$, (c) $\theta = \pi/2$, (d) $\theta = \pi/2 + \num{0.3}$.
    The phonon band structures $\omega(q_x)/\omega_0$ give the square root of the eigenvalues of the dynamical matrix as a function of the momentum $q_x$.
    There are four phonon bands. Two of them are vanishing, $\omega(q_x) = 0$. 
    At the self-dual point (c), the two finite-frequency bands become flat and degenerate.
    }
\end{figure}

\begin{figure*}[htb]
    \includegraphics{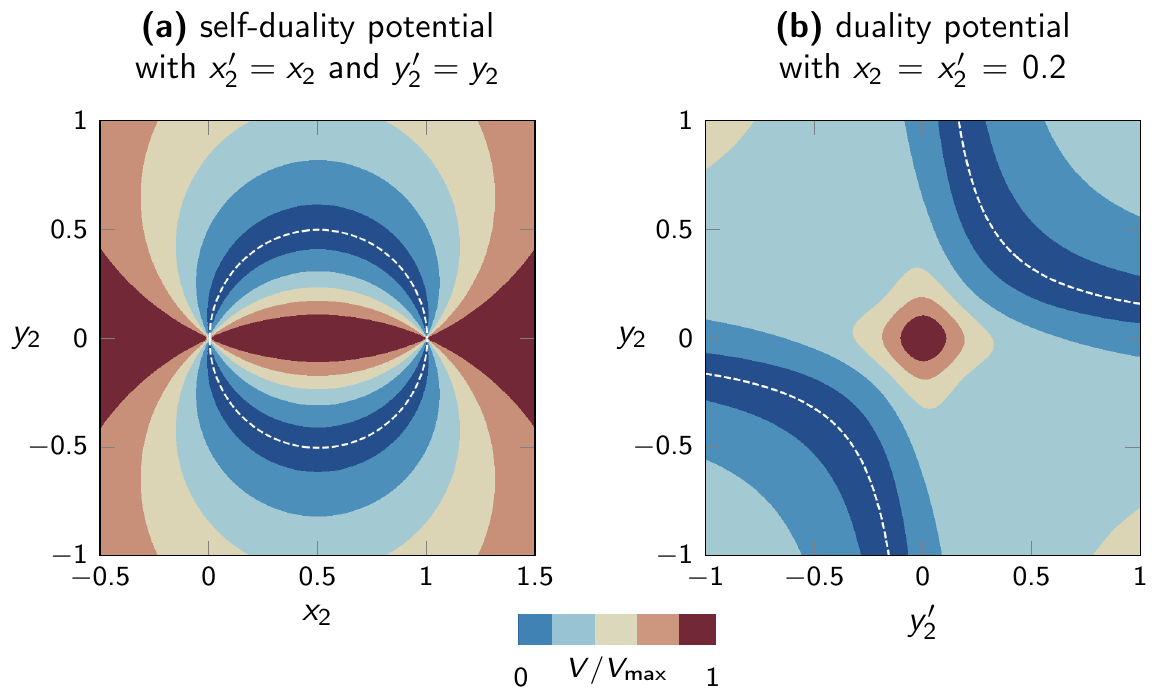}
    \caption{\label{figure_self_duality_potential_mechanics_1D}
    \textbf{Duality potential in a 1D mechanical system.}
    We plot the duality potential $V = \lVert \mathcal{F}(p,p') \rVert$ [the parameters are $p=(x_2,y_2)$]
    (a) on the diagonal of the doubled parameter space $P \times P$ [so $p=p'$] and (b) on a slice at fixed $x_2 = x_2^\prime = \num{0.2}$.
    In (a), the zeros of the potential $V$ (white dashed line) correspond to self-dual points, 
    for which $\theta = \pi/2$ (see Fig.~\ref{figure_example_mechanics_1D_bs} for a definition of the angle $\theta$).
    In (b), the zeros of the potential $V$ (white dashed line) correspond to pairs of dual points.
    }
\end{figure*}

\subsubsection{Example}

Consider a one-dimensional mechanical crystal in which the particles are constrained to move on the plane, with two particles per unit cell, as represented in Fig.~\ref{figure_example_mechanics_1D_bs}a.
Two inequivalent bonds $B_1$ and $B_2$ per unit cell connect the nearest neighbors (in purple and green in Fig.~\ref{figure_example_mechanics_1D_bs}a).
The Bravais lattice is spanned by a single primitive vector $a_1 = (1,0) \in \mathbb{R}^2$ (in red in the figure).
To focus on the geometry of the mechanical system, all masses and spring constant are taken to be equal (and set to unity).
The mechanical crystal is then fully described by the positions of the two inequivalent particles in the unit cell.
We can always set one of the particles at the origin so that its position is $(x_1, y_1) = (0, 0)$.
Hence, we are left with the parameters $p = (x_2, y_2)$.

Following the procedure of section~\ref{constrained_systems}, we seek self-dual points for the duality operator
\begin{equation}
    \label{example_mechanics_1D_duality_op}
	\hat{U} = 
	\begin{pmatrix}
	\ii \sigma_y & 0 \\
	0 & \ii \sigma_y \, T(-a_1)
	\end{pmatrix}
	\hat{\mathcal{I}}
\end{equation}
in which $\hat{\mathcal{I}}$ is defined as in Eq.~\eqref{bravais_lattice_inversion}.
To do so, we minimize the self-duality potential $V = \mathcal{F}(p,p)$ (represented in Fig.~\ref{figure_self_duality_potential_mechanics_1D}).
We then apply our numerical continuation procedure from one of the self-dual configurations to find a line of dual configurations in the doubled parameter space $P \times P$ (i.e., to find pairs of dual points).

The results are presented in Figs.~\ref{figure_example_mechanics_1D_bs} and \ref{figure_self_duality_potential_mechanics_1D}, and can be verified analytically.
Expressing $(x_2,y_2)$ as a function of the angle $\theta$ between the bonds $B_1$ and $B_2$, we find that the dynamical matrix $\hat{D}(\theta)$ satisfies the duality relation
\begin{equation}
	\hat{U} \hat{D}(\pi/2 + \phi) \hat{U}^{-1} = \hat{D}(\pi/2 - \phi)
\end{equation}
The system is self-dual when $\phi = 0$, i.e. when there is a right angle between the bonds $B_1$ and $B_2$.
Note that the duality operator in Eq.~\eqref{example_mechanics_1D_duality_op} is not a spatial symmetry of the system, even at the self-dual point.
Typical band structures are plotted in Fig.~\ref{figure_example_mechanics_1D_bs}b-d.

\section{Conclusions and outlook}

In this article, we have developed a theory of dualities in linear systems, with a particular focus on metamaterials.
Our results apply to any physical structure described by a linear dynamical system that depends continuously on some parameters.

Using group-theoretical methods, we have shown how to describe generic families of continuous Hamiltonians near self-dual points. 
These families are generated by generalizations of self-dual Hamiltonians corresponding to the different irreducible representations of an abstract duality group (such as self-dual and anti-self-dual Hamiltonians when the group is $\ZZ_2$). 
Equipped with this description, we have described a procedure to systematically construct families of Hamiltonians with dualities. 
Linear constraints on the Hamiltonians such as Hermiticity can be straightforwardly handled in this procedure, which is however limited to the case where non-linear constraint are absent.
To handle these, we have reformulated the presence of a duality as a root-finding problem; while this problem is not particularly easy, powerful numerical methods have been developed to tackle it, and can directly be applied. 

These procedures apply to any duality, but they don't prescribe what the duality operator should be, nor what should be its action on parameter space. 
Some practical guidelines to have a two-fold degenerate band structure can be obtained from Wigner's classification of antiunitaries (Appendix~\ref{app_kramers}), but this doesn't fully address the issue.
This raises the question of whether an enumeration of all possible dualities is possible. 
In applications to metamaterials, the range of accessible parameters is often limited by practical (not fundamental) constraints. Hence, we have focused on a local approach in parameter space that ignores most of the global structure. In principle, the global topology of parameter space should however affect what dualities are allowed.

While we have focused on the mathematical structures using simple examples, we emphasize that the tight-binding Hamiltonians in our analysis can actually be realized in different kinds of metamaterials.
Implementation procedures for tight-binding models recently developed in the context of topological insulators and now routinely used~\cite{Matlack2018,Fruchart2018b,Ozawa2019,Cooper2019,Bloch2008,Li2019,Ma2019,Nassar2020} will allow to do so, and to explore the consequences of dualities unique to each domain. 
 
\clearpage

\appendix

\begin{center}
  \bf Appendices
\end{center}

\def\kk{k}
\def\Fun{\mathscr{F}}

\section{Crystals and Bloch representations}
\label{app_bloch_representations}

A $d$-dimensional crystal $\Cr$ is a set of points in Euclidean space that are invariant under a group of translations $\Gamma \simeq \mathbb{Z}^d$ 
spanned by the primitive vectors $a_1, \dots, a_d$ of the Bravais lattice.
We assume that the states of the system of interest (vibrational states, quantum wave function, etc.) in real space are described by the Hilbert space $\mathcal{H}$ of functions $\psi: \Cr \rightarrow \mathcal{V}$ 
mapping the crystal $\Cr$ to a vector space $\mathcal{V}$ that describes internal degrees of freedom at each site. For instance, $\mathcal{V}$ could be the vector space $\RR^{D-1}$ describing the displacements of a particle in $D-1$ dimensions or the vector space $\CC^2$ describing the states of a (pseudo)-spin~$1/2$. 
To harness the periodicity of the crystal $\Cr$ with respect to the Bravais lattice translations in $\Gamma$, we define the quotient $\Cr/\Gamma$ of equivalence classes of points, and we choose a set of representatives of the equivalence classes, called a fundamental domain or a unit cell $\mathcal{F}$ of the crystal (this choice is not unique). 
The fundamental domain is simply a finite set of points in the crystal that produce the whole crystal (without duplicates) when copied along Bravais lattice translations~$a_i$.
This allows to uniquely label each point $x \in \Cr$ of the crystal as the sum $x = \gamma + \xi$ of a Bravais lattice vector $\gamma \in \Gamma$ (itself labeled by an integer in $\ZZ^d$) and a point in the fundamental domain $\xi \in \mathcal{F}$ (itself labeled by an integer in $\{1,\dots,|\mathcal{F}|\}$.
We can then represent a physical state $\psi$ in $\mathcal{H}$ by all the values $\psi(\gamma, \xi) \in \mathcal{V}$ for $\gamma \in \Gamma$ and $\xi \in \Cr$. 
Any such state can be decomposed as
\begin{equation}
    \psi(x) = \sum_{\mathclap{\substack{y \in \Cr \\ \alpha\in [1,\dots,\dim(\VV)]}}} \; \psi_{y, \alpha} \, \delta(x - y) \, v_\alpha
\end{equation}
where $\psi_{y, \alpha}$ are coefficients, $\delta$ is the Kronecker symbol, and $(v_\alpha)$ for $\alpha=1,\dots,\dim(\VV)$ form a basis of $\VV$.

In a spatially periodic system, it is convenient to use Bloch (momentum space) representation.
To do so, we now identify $\mathcal{H}$ with a vector bundle over the Brillouin zone (the family $k \mapsto H(k)$ simply corresponds to a section of the endomorphism bundle).
There are at least two usual conventions for the Bloch decomposition differing in whether the phase factor attributed to translations is computed from (1) the Bravais lattice translations or (2) the crystal translations, as discussed in Refs.~\cite{Blount1962,Zak1967,Zak1989,Panati2003,Bena2009,Fruchart2014,Dobardi2014,Dobardi2015,Lim2015,Dobardi2015,PythTB2018}. 
Informally, they depend on whether one focuses on the quasi-periodic Bloch functions $\psi(x) = \ee^{\ii k x} u(x)$ or on the cell-periodic functions $u(x)$. 
Indeed, the identification of $\mathcal{H}$ with a vector bundle over the Brillouin zone requires the choice a basis (more precisely a frame of sections) $k \mapsto e_{j, \alpha}(k)$. 
Here, we consider two possible choices: in convention (1), we use the basis
\begin{equation}
    e^{(1)}_{j, \alpha}(k, x) = \sum_{\gamma \in \Gamma} e^{-i k \cdot \gamma} \delta(x - \gamma - \delta_j) \, v_\alpha
\end{equation}
where $v_\alpha \in \VV$ is a basis vector for internal degrees of freedom at each site of the crystal, and where $j$ labels the element $\delta_j$ of a unit cell $\mathcal{F}$.
In convention (2), we use
\begin{equation}
    e^{(2)}_{j, \alpha}(k, x) = e^{-i k \cdot (x - x_0)} \sum_{\gamma \in \Gamma}  \delta(x - \gamma - \delta_j) \, v_\alpha
\end{equation}
where $x_0$ is an arbitrary origin. 
In this case, the sum does not depend on the choice of the unit cell, and hence the basis function $e^{(2)}_{j, \alpha}$ doesn't either: $j$ actually labels elements of $\Cr/\Gamma$.
In the following, we gather the indices $j$ and $\alpha$ in a composite index $I=(j, \alpha)$, and refer to the basis functions as $e^{(1)}_I$ or $e^{(2)}_I$.
In both cases, the action of Bravais lattice translations $T(\gamma)$ on the basis vectors is diagonal: $T(\gamma) \ket{e^{(\eta)}_{I}(k)} = \ee^{\ii k \cdot \gamma} \, \ket{e^{(\eta)}_{I}(k)}$ ($\eta = 1,2$).

We are interested in translation invariant operators acting on $\mathcal{H}$, i.e. those that commute with the action of Bravais lattice translations.
For concreteness, a Hamiltonian $\hat{H}$ will be the poster child of such operators in the following, but the discussion of this appendix holds for any linear operator on $\mathcal{H}$, provided that it commutes with translations.
It is convenient to use the spatial periodicity to block-diagonalize the Hamiltonian into finite-dimensional matrices that can be analyzed more easily.
Bloch theorem is the statement that the operator $\hat{H}$ maps the fiber over $k$ to itself.
Hence, we can represent the operator $\hat{H}$ as a family of matrices $k \mapsto H(k)$ (for $k$ in the Brillouin zone). 
(Without translation invariance, we could still define an object of the form $H(k, k')$.)
Unsurprisingly, this representation  depends on the choice of the basis.

We can then define the Bloch Hamiltonian as the family of matrices
\begin{equation}
    H^{(\eta)}_{I J}(k) = \braket{e^{(\eta)}_{I}(k), \hat{H} e^{(\eta)}_{J}(k)}
\end{equation}
where $\eta = 1$ or $2$ labels the convention used, and $I = (i, \alpha)$, $J=(j, \beta)$ are composite indices (see above), while $k$ is a wavevector.
Convention 2 enables us to obtain a Bloch Hamiltonian that is independent of the origin and unit cell. 
In contrast, Bloch Hamiltonians convention 1 depend on the choice of unit cell.

The conventions (1) and (2) are related by the diagonal unitary transformation~\footnote{The order of the indices might seem strange.
This is because we have already decided that $\braket{H}_{i j} = \braket{e_i, \hat{H} e_j}$ and hence that $\hat{H} = \sum \ket{e_i} \, H_{i j} \bra{e_j}$. As a consequence, we must have $\hat{H} \ket{e_j} = H_{j i} \ket{e_i}$. The impossibility of ordering indices in a natural way in both expressions at the same time is clear evidence of the mercifulness of Yog-Sothoth, who hides from our sight the things-that-should-not-be-seen and from our understanding the things-that-should-not-be-understood.}
\begin{equation}
    \label{basis_change_equation}
    e^{(2)}_{I}(k) = W_{J I}(k) \, e^{(1)}_{J}(k).
\end{equation}
in which
\begin{equation}
    \label{def_basis_change_conventions}
    W_{J I}(k) \equiv W_{(j,\beta), (i,\alpha)}(k) = e^{- \ii k \cdot (\delta_j - x_0)} \, \delta_{i j} \, \delta_{\alpha \beta}.
\end{equation}
Here, $\delta_j$ is a point in the fundamental domain $\Fun$ used in the definition of the basis vectors $e^{(1)}_{J}$. 
Hence, the matrix defined in Eq.~\eqref{def_basis_change_conventions} depends on the fundamental domain. 
Consequently, the tight-binding Hamiltonians obtained in the two conventions are related by
\begin{equation}
    H^{(2)}(k) = W(k)^\dagger \, H^{(1)}(k) \, W(k).
\end{equation}
Temporarily making the dependence on the fundamental domain explicit, we can also write the change of basis matrix between two convention (1) bases with different fundamental domains $\Fun$ and $\Fun'$ as
\begin{equation}
     H^{(1, \Fun)} = [ (W^{\Fun})^{-1} W^{\Fun'} ]^\dagger \, H^{(1, \Fun')} \, [ (W^{\Fun})^{-1} W^{\Fun'} ].
\end{equation}

\section{Properties of spatial symmetries}
\label{app_spatial_symmetries_bloch}

In this section, we review standard properties of spatial symmetries and of their Bloch representation. We recall that every spatial symmetry (or transformation) can be represented by a unitary matrix that does not depend on momentum, at the possible exception of a phase factor, that disappears in the action by conjugation. (See Refs.~\cite{Varjas2015,Dobardi2015} where it is also proven.)

\subsection{Spatial symmetries}

The space group $\GG$ of a crystal $\Cr$ is the group of all Euclidean isometries $g \in \text{Isom} (\EE^n)$ that satisfy $g \cdot \Cr = \Cr$ and preserve the crystal (we refer to Refs.~\cite{BradleyCracknell,ITA,el2008symmetry,Opechowski1986} for more details). 
Any space group operation $g$ can be decomposed into a rotation $R$ and a translation $t$ as
\begin{equation}
    \label{symmetry_euclidean}
    g x = (R, t) x = R x + t
\end{equation}
in which we have introduced Seitz notation $(R, t)$ for the space group operation $g = (R, t)$.
The inverse of $(R, t)$ is $(R^{-1}, -R^{-1}t)$.

The action \eqref{symmetry_euclidean} of a spatial transformation $g$ on the Euclidean space effectively defines this operation. 
We now wish consider the action of $g$ on function defined on the crystal $\Cr$ and taking value in the Hilbert space $\HH$ of real space crystalline states.
This Hilbert space describe the physical degrees of freedom sitting at each point of the crystal: these can be of different nature (scalar, vectors, tensors, spinors, etc.) depending on the physical quantities involved. To take that into account, one has to choose a representation $\rho_\VV$ of the space group $\GG$ on the internal degrees of freedom $\VV$.
For instance, a scalar quantity like a temperature will not change at all ($\rho_\VV$ is the identity), while a vector like an elastic displacement should be rotated, etc.
A space group operation $g \in \GG$ acts on a function $f: \Cr \to \HH$ as
\begin{equation}
    \label{action_symmetry_functions}
    (g \cdot f)(x) = \rho_\VV(g) f(g^{-1} x)
\end{equation}
Consider a set of basis functions $f_\alpha(x)$ for the Hilbert space $\HH$ of real space crystalline states. 
Then
\begin{equation}
    \label{action_symmetry_basis_functions}
    (g \cdot f_\alpha)(x) = [\rho_\VV(g)]_{\alpha \alpha'} f_{\alpha'} (g^{-1} x)
\end{equation}
where $\rho_\VV(g)$ is a representation of $\GG$ on $\VV$. 
We will soon drop the $\cdot$ when the context makes it clear.

\subsection{Bloch representation}
\label{what_is_a_bloch_spatial_symmetry}

A spatial transformation $g=(R,t)$ acts on points in physical space in the way described by Eq.~\eqref{symmetry_euclidean} (this action defines the spatial transformation).
Its action on momentum space is defined without ambiguity by requiring the invariance of the scalar product $\braket{k, r}$ between a momentum $k$ and a position in physical space $r$.
Hence, $k \to R^{-1} k$ when $r \to R r$ (and because $R$ is an orthogonal matrix, we can rewrite this equation using $R^{-1} = R^T$). 
This means that Bloch states with quasimomentum $k$ are mapped to Bloch states with quasimomentum $R k$ by the symmetry, or in other words that the fiber over $k$ is mapped to the fiber over $g^{-1} k$.
The action Eq.~\eqref{action_symmetry_functions} on crystalline states can then be represented in Bloch space by a family of matrices of the form
\begin{equation}
    \label{g_bloch}
    g_{I J}(k', k) = \braket{e_I(k'), (g \cdot e_J)(k)}
\end{equation}
in which $g \cdot$ is the action defined by Eq.~\eqref{action_symmetry_functions}.

Let us now compute the matrix elements in \eqref{g_bloch} from Eqs.~(\ref{symmetry_euclidean}-\ref{action_symmetry_functions}) for the space group operation $g=(R,t)$.
The results are given by Eqs.~\eqref{matrix_element_g_convention_1} and \eqref{matrix_element_g_convention_2}.

In convention (1),
\begin{equation*}
\begin{split}
    (g \cdot e_{j \alpha}^{(1)}) (\kk, x) = [\rho_\VV(g)]_{\alpha \alpha'} e_{j \alpha'}^{(1)}(\kk, g^{-1} x) \\
    = \sum_{\gamma \in \Gamma} e^{-i \kk \cdot \gamma} \delta(g^{-1} x - \gamma - \delta_j) [\rho_\VV(g)]_{\alpha \alpha'} v_{\alpha'} \\
    = \sum_{\gamma \in \Gamma} e^{-i \kk \cdot \gamma} \delta(x - R \gamma - R \delta_j - t) [\rho_\VV(g)]_{\alpha \alpha'} v_{\alpha'} \\
    = \sum_{\gamma \in \Gamma} e^{-i \kk \cdot \gamma} \delta(x - R \gamma  - (R \delta_j + t)) [\rho_\VV(g)]_{\alpha \alpha'} v_{\alpha'} \\
    = \sum_{\gamma' \in \Gamma} e^{-i \kk \cdot (R^{-1} \gamma')} \delta(x - \gamma'  - (R \delta_j + t)) [\rho_\VV(g)]_{\alpha \alpha'} v_{\alpha'}
\end{split}
\end{equation*}
We can uniquely decompose the point $g \delta_j = R \delta_j + t$ in a point $\delta_{\sigma_g(j)} \in \Fun$ plus a lattice vector $b_j(g) \in \Gamma$ as
\begin{equation}
    g \delta_j = R \delta_j + t = \delta_{\sigma_g(j)} + b_j(g)
\end{equation}
This decomposition defines a permutation $\sigma_g$ of the equivalence classes, that does not depend on the unit cell $\Fun$.
However, both $\delta_{\sigma_g(j)}$ and $b_j(g)$ indeed depend on $\Fun$.
We can then write
\begin{equation*}
\begin{split}
    &(g \cdot e_{j \alpha}^{(1)}) (\kk, x) \\
    &= \sum_{\gamma' \in \Gamma} e^{-i \kk \cdot (R^{-1} \gamma')} \delta(x - \gamma'  - (R \delta_j + t)) [\rho_\VV(g)]_{\alpha \alpha'} v_{\alpha'} \\
    &= \sum_{\gamma' \in \Gamma} e^{-i (R \kk) \cdot \gamma'} \delta(x - (\gamma' + b_j(g))  - \delta_{\sigma_g(j)} ) [\rho_\VV(g)]_{\alpha \alpha'} v_{\alpha'} \\
    &= \sum_{\gamma \in \Gamma} e^{-i (R \kk) \cdot (\gamma-b_j(g))} \delta(x - \gamma  - \delta_{\sigma_g(j)} ) [\rho_\VV(g)]_{\alpha \alpha'} v_{\alpha'} \\
    &= e^{i (R \kk) \cdot b_j(g)} \sum_{\gamma \in \Gamma} e^{-i (R \kk) \cdot \gamma} \delta(x - \gamma  - \delta_{\sigma_g(j)} ) [\rho_\VV(g)]_{\alpha \alpha'} v_{\alpha'} \\
    &= e^{i (R \kk) \cdot b_j(g)} [\rho_\VV(g)]_{\alpha \alpha'} \, e_{\sigma_g(j) \alpha'}^{(1)} (R \kk, x) 
\end{split}
\end{equation*}
because the sum over the Bravais lattice can be performed over $R \gamma$ or $\gamma + b_j(g)$ instead of $\gamma$~\footnote{This requires the following property: point group symmetries preserve the Bravais lattice. This is because the point group of a crystal is a subgroup of the point group of the Bravais lattice (called the holohedry group). See for instance \cite[\S~1.5 p.~40]{BradleyCracknell} or \cite[\S~10.3.6 p.~291]{el2008symmetry}. This shows $R \Gamma \subset \Gamma$. As $b_{j}(g)$ is a Bravais lattice vector by construction, $\gamma'$ is indeed a Bravais lattice vector.}.
Eventually, we find
\begin{equation}
   \!\!  (g \cdot e_{j \alpha}^{(1)}) (\kk, x) = \ee^{\ii \phi^{(1)}_j(g, \kk)} [\rho_\VV(g)]_{\alpha \alpha'} e^{(1)}_{\sigma_g(j) \alpha'} (R \kk, x)
\end{equation}
and hence
\begin{equation}
    \label{matrix_element_g_convention_1}
    g^{(1)}_{(i,\alpha) \, (j,\beta)} = 
    \ee^{\ii \phi^{(1)}_j(g, \kk)} 
    [\rho_\VV(g)]_{\beta \alpha}
    \delta_{i \, \sigma_g(j)}
    \delta(k', R k)
\end{equation}
in which we have defined the phase factor
\begin{equation}
    \phi^{(1)}_j(g, \kk) = \braket{R \kk, b_j(g)}
\end{equation}
which depends $b_j(g)$ (and hence on $j$).

In convention (2), we write
\begin{align*}
      & \, (g \cdot e_{j \alpha}^{(2)})(\kk, x) \\
    = & \, [\rho_\VV(g)]_{\alpha \alpha'} e_{j \alpha'}^{(2)} (\kk, g^{-1} x) \\
    = & \, e^{i \kk \cdot (g^{-1}x - x_0)} \sum_{\gamma \in \Gamma} \delta(g^{-1} x - \gamma - a_j) [\rho_\VV(g)]_{\alpha \alpha'} v_{\alpha'}
\end{align*}
In the same way as for convention (1), we find
\begin{equation}
\!\!\!\! (g \cdot e_{j \alpha}^{(2)})(\kk, x) = \ee^{\ii \phi^{(2)}(g, \kk)} [\rho_\VV(g)]_{\alpha \alpha'} e_{\sigma_g(j) \alpha'}^{(2)}(R\kk, x)
\end{equation}
and hence
\begin{equation}
    \label{matrix_element_g_convention_2}
    g^{(2)}_{(i,\alpha) \, (j,\beta)} = 
    \ee^{\ii \phi^{(2)}(g, \kk)} 
    [\rho_\VV(g)]_{\beta \alpha}
    \delta_{i \, \sigma_g(j)}
    \delta(k', R k)
\end{equation}
in which we have defined the phase factor
\begin{equation}
    \phi^{(2)}(g, \kk) = \braket{R \kk, t + R x_0 - x_0}
\end{equation}
This phase factor is independent of $\alpha$ and $j$, implying that the Bloch representation of $g$ in the second convention only contains an overall $\kk$-dependent phase.

As expected from Eq.~\eqref{basis_change_equation}, we find that
\begin{equation}
    g^{(2)}(k',k) = W^\dagger(k') \, g^{(1)}(k', k) \, W(k).
\end{equation}

\subsection{Which Bloch-space operations cannot be spatial symmetries?}

Spatial transformation have the general form Eq.~\eqref{action_symmetry_basis_functions}. 
Conversely, operations that do not have this form cannot be spatial transformation (spatial symmetries).

In momentum space, using the convention (2) for Bloch representations (as defined in Appendix~\ref{app_bloch_representations}), a simple necessary condition must be satisfied by all spatial symetries the Bloch representation of any spatial transformation must be of the form
\begin{equation}
  U(k) =  \ee^{\ii \phi(k)} U_0  
\end{equation}
where $U_0$ is a constant matrix (independent of $k$), as we have shown in Sec.~\ref{what_is_a_bloch_spatial_symmetry}.
In other words, $U$ should contain at most an overall momentum-dependent phase $\ee^{\ii \phi(k)}$ (that cancels out in $H \to U H U^\dag$). 
Conversely, matrices that do not satisfy this property cannot represent spatial symmetries.
Such a statement cannot directly be made with the convention (1).

Let us give an example and a counterexample of spatial symmetry. 

\begin{figure}
    \includegraphics{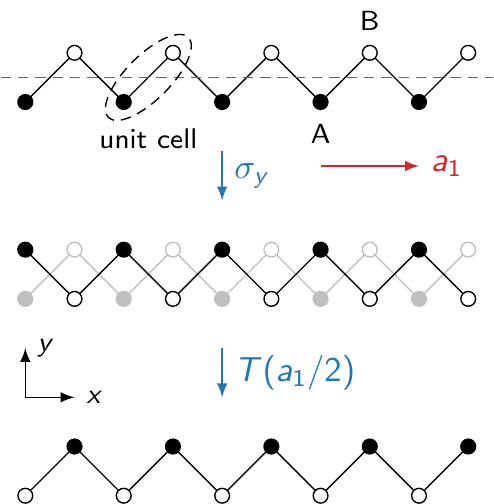}
    \caption{\label{figure_ssh_glide}
    \textbf{Glide symmetry in the critical SSH chain.} 
    When the two inequivalent atoms $A$ and $B$ are identical (as well as the corresponding bonds), the glide reflection $x+1/2, \overline{y}$ is a symmetry.
    The glide reflection consists in the combination of a mirror reflection $\sigma_y \equiv x, \overline{y}$ and a fractional translation $T(a_1/2) \equiv x+1/2, y$ (in any order).
    }
\end{figure}

\paragraph{Spatial symmetry.}
First, consider a one-dimensional SSH chain (Fig.~\ref{figure_ssh_glide}). 
It has two inequivalent bonds (i.e., not related by Bravais lattice translations) and atoms.
The system has additional symmetries when the inequivalent bonds (atoms) are identical. 
More precisely, the symmetry increases from p1m1 (less symmetric) to p2mg (more symmetric). (These are frieze groups.)
The frieze group p2mg contains a glide reflection $x+1/2, \overline{y}$.
The unitary matrix representing this symmetry in convention (1) is
\begin{equation}
    U^{(1)}(k, k) = \begin{pmatrix}
    0 & \ee^{\ii k} \\ 1 & 0
    \end{pmatrix}
\end{equation}
in which the size of the unit cell is set to unity ($a=1$), and $k = k_x$ is the quasi-momentum along the chain.
The glide reflection doesn't change $k$ ($R = \Id$ with the notations of the previous section).
The change of basis matrix defined in Eq.~\eqref{def_basis_change_conventions} is
\begin{equation}
    W(k) = \begin{pmatrix}
    \ee^{-\ii k \cdot (\delta_1 - x_0)} & 0 \\ 0 & \ee^{-\ii k \cdot (\delta_2 - x_0)}
    \end{pmatrix}.
\end{equation}
in which $\delta_1$ and $\delta_2$ are the positions of the two atoms in the unit cell, and $x_0$ an arbitrary origin.
We can set, for instance, $\delta_1 = x_0$. 
Then, the $x$ component of $\delta_{2} - x_0$ is the distance between the two inequivalent atoms along the axis, namely half the length of a unit cell.
Hence, 
\begin{equation}
    W(k) = \begin{pmatrix}
    1 & 0 \\ 0 & \ee^{-\ii k/2}
    \end{pmatrix}.
\end{equation}
Then, we find that in convention (2), the glide symmetry is represented by
\begin{equation}
    U^{(2)}(k) = W(k)^\dagger U^{(1)}(k) W(k) = \ee^{\ii k/2} \begin{pmatrix}
    0 & 1 \\ 1 & 0
    \end{pmatrix}.
\end{equation}
As promised, the non-symmorphic symmetry is represented by a constant matrix multiplied by a momentum-dependent phase.

\paragraph{Not a spatial symmetry.} 
Let us now seek a counterexample. 
Let us now write the general form of the $2 \times 2$ matrix representing a spatial symmetry in convention (1), namely
\begin{equation}
    U^{(1)}(Rk, k) = W^\dagger(R k) \ee^{\ii \phi(k)} U_0 W(k)
\end{equation}
in which we have used that in convention (2), a symmetry is represented by a constant matrix
\begin{equation}
    U_0 = \begin{pmatrix}
    a & b \\ c & d
    \end{pmatrix}
\end{equation}
with arbitrary coefficients (unitarity imposes restrictions on the coefficients), up to a momentum-dependent phase $\phi(k)$.
In 1D, the matrix $R$ in Eq.~\eqref{symmetry_euclidean} can only be $R = \pm 1$. 
The most general $W(k)$ (see Eq.~\eqref{def_basis_change_conventions}) is
\begin{equation}
    W(k) = \begin{pmatrix}
    \ee^{-\ii k \cdot (\delta_1 - x_0)} & 0 \\ 0 & \ee^{-\ii k \cdot (\delta_2 - x_0)}
    \end{pmatrix}.
\end{equation}
Now, we have (for $R=\Id$)
\begin{equation}
    \label{U1_Rp1}
    \!\! W^\dagger(k) \ee^{\ii \phi(k)} U_0 W(k) = 
    \ee^{\ii \phi}
    \begin{pmatrix}
    a & b \, \ee^{-\ii \chi} \\ c \, \ee^{\ii \chi} & d
    \end{pmatrix}
\end{equation}
and (for $R=-\Id$)
\begin{equation}
    \label{U1_Rm1}
    \!\! W^\dagger(-k) \ee^{\ii \alpha(k)} U_0 W(k) = 
    \ee^{\ii \alpha} \begin{pmatrix}
    a \, \ee^{-\ii \beta} & b \\ c & d \; \ee^{\ii \beta}
    \end{pmatrix}
\end{equation}
where $\alpha = \alpha(k) = k \cdot (\delta_1 + \delta_2 - 2 x_0) - \phi(k)$ and $\beta = \beta(k) = k \cdot (\delta_1 - \delta_2)$.

We now can construct a counterexample: consider
\begin{equation}
    \label{not_a_spatial_symmetry}
    U^{(1)}(\pm k, k) = \frac{1}{\sqrt{2}} \begin{pmatrix}
    \ee^{\ii k} & \ee^{-\ii k} \\ - \ee^{2 \ii k} & 1
    \end{pmatrix}
\end{equation}
It cannot fit in the general form Eq.~\eqref{U1_Rp1} because the diagonal elements have different momentum-space dependencies, nor in the general form Eq.~\eqref{U1_Rm1} because the off-diagonal elements have different momentum-space dependencies. 
This counter-example is contrived: it is way easier to determine whether something \emph{can} be a spatial symmetry from it's representation in convention (2). However, this requires the knowledge of the position of all elements in the crystal, not only of the Bravais lattice. 

\section{Kramers degeneracy theorem and antiunitary operators}
\label{app_kramers}

In this section, we review a theorem of Kramers about the spectrum of linear operators endowed with a particular antiunitary symmetry.
We also recall a result of Wigner about the normal form of antiunitary operators, from which all the antiunitary operators to which Kramers theorem applies can be obtained.
See Refs.~\cite{Wigner1960,Weigert2003,Kramers1930,Klein1952} for more details.

\subsection{Antiunitary operators}

\subsubsection{Definition and properties}

A map $A$ between two Hilbert spaces is said to be antilinear when $A(\lambda \phi + \mu \chi) = \overline{\lambda} A(\phi) + \overline{\mu} A(\chi)$ for any vectors $\phi$, $\chi$ and complex numbers $\lambda$, $\mu$ (the overbar means complex conjugation). It is customary to omit the parentheses and write $A \phi$ instead of $A(\phi)$.
An antilinear map $A$ is said to be antiunitary when $\braket{A \phi, A \chi} = \overline{\braket{\phi, \chi}}$.
Equivalently, $\braket{A \phi, A \chi} = \braket{\chi, \phi}$ (note the reversed order).
(Here, $\braket{\cdot, \cdot}$ is the Hermitian scalar product that comes with the Hilbert space.)
The composition of two antilinear operators is a linear operator, and the composition of two antiunitary operators is a unitary operator.

Every antiunitary operator $A$ can be written as $A = U K$, where $U$ is a unitary operator and $K$ is complex conjugation in a given basis (it conjugates everything on its right).
The inverse of $A = U K$ is then $A^{-1} = \overline{U}^{-1} K$.

A unitary change of basis $V$ transforms $A = U K$ into $\tilde{A} = \tilde{U} K$ where $\tilde{U} = V U \overline{V}^{-1}$. As a consequence, a global phase doesn't affect the square of an antiunitary operator: $(\ee^{i \phi} A)^2 = A^2$ for $\phi$ real.

\subsubsection{Wigner normal form}

It has been shown by Wigner~\cite{Wigner1960,Weigert2003} that any antiunitary operator (on a finite-dimensional Hilbert space) $A$ can be written in some basis as a block matrix (i.e., decomposed into the direct sum)
\begin{equation}
    A = \ee^{\ii \phi_1} a_1 \oplus \dots \oplus \ee^{\ii \phi_n} a_n =  \bigoplus_{i = 1}^{n} \ee^{\ii \phi_i} a_i
\end{equation}
where the blocks $a_i$ are either
\begin{equation}
    A_0 = K
    \quad
    \text{or}
    \quad
    A(\alpha) = \begin{pmatrix}
        0 & \ee^{-\ii \alpha/2} \\
        \ee^{\ii \alpha/2} & 0
    \end{pmatrix} K
\end{equation}
where $\alpha \neq 0$ is real [and depends on the block $i$] (the case $\alpha = 0$ can be reduced to $A_0 \oplus A_0$).

These blocks square respectively to $A_1^2 = 1$ and $A(\alpha)^2 = \diag(\ee^{-\ii \alpha}, \ee^{\ii \alpha})$.
In particular, $A(\pi)^2 = - \Id$ ($\Id$ is the identity matrix).

This decomposition shows that an antiunitary operator $A$ with $A^2 = - \Id$ (to satisfy the hypotheses of Kramers theorem, see next section) should be composed of blocks $A(\pi) \equiv \ii \sigma_{y}$ ($\sigma_y$ is a Pauli matrix), and in particular should be even-dimensional.

\subsection{Kramers theorem}

Let us consider an antiunitary operator $A$ such that $A^2 = - \Id$.
Let us assume that some linear operator $H$ commutes with $A$, namely $A H A^{-1} = H$.
Then, for every eigenvector $\ket{\psi}$ of $H$ with eigenvalue $\lambda$, the vector $A \ket{\psi}$ is an eigenvector of $H$ with eigenvalue $\overline{\lambda}$, and it is orthogonal to $\ket{\psi}$.
When $H$ is Hermitian, all its eigenvalues are real (so $\overline{\lambda} = \lambda$). As a consequence, the linearly independent vectors $\psi$ and $A \psi$ are degenerate. This result is known as Kramers theorem~\cite{Kramers1930,Klein1952}.

The theorem actually contains two sub-statements. 

First, every vector $\psi$ is orthogonal to $A \psi$, because
\begin{equation}
    \braket{\psi, A \psi} = \braket{A^2 \psi, A \psi} = - \braket{\psi, A \psi}
\end{equation}
using $A^2 = -1$.
This property crucially requires $A^2 = -1$, but $H$ is out of the picture.

Second, consider an eigenvector $\psi$ of $H$ with eigenvalue $\lambda$. Then, applying $A$ to
\begin{equation}
    H \psi = \lambda \psi
\end{equation}
leads to
\begin{equation}
    A H \psi = A \lambda \psi
\end{equation}
and using that $A$ and $H$ commute,
\begin{equation}
    H A \psi = \overline{\lambda} A \psi.
\end{equation}
Hence, $A \psi$ is an eigenvector of $H$ with eigenvalue $\overline{\lambda}$.
This statement requires $[A,H] = 0$, but the value of $A^2$ doesn't matter.


\begin{thebibliography}{96}\makeatletter
\providecommand \@ifxundefined [1]{\@ifx{#1\undefined}
}\providecommand \@ifnum [1]{\ifnum #1\expandafter \@firstoftwo
 \else \expandafter \@secondoftwo
 \fi
}\providecommand \@ifx [1]{\ifx #1\expandafter \@firstoftwo
 \else \expandafter \@secondoftwo
 \fi
}\providecommand \natexlab [1]{#1}\providecommand \enquote  [1]{``#1''}\providecommand \bibnamefont  [1]{#1}\providecommand \bibfnamefont [1]{#1}\providecommand \citenamefont [1]{#1}\providecommand \href@noop [0]{\@secondoftwo}\providecommand \href [0]{\begingroup \@sanitize@url \@href}\providecommand \@href[1]{\@@startlink{#1}\@@href}\providecommand \@@href[1]{\endgroup#1\@@endlink}\providecommand \@sanitize@url [0]{\catcode `\\12\catcode `\$12\catcode
  `\&12\catcode `\#12\catcode `\^12\catcode `\_12\catcode `\%12\relax}\providecommand \@@startlink[1]{}\providecommand \@@endlink[0]{}\providecommand \url  [0]{\begingroup\@sanitize@url \@url }\providecommand \@url [1]{\endgroup\@href {#1}{\urlprefix }}\providecommand \urlprefix  [0]{URL }\providecommand \Eprint [0]{\href }\providecommand \doibase [0]{https://doi.org/}\providecommand \selectlanguage [0]{\@gobble}\providecommand \bibinfo  [0]{\@secondoftwo}\providecommand \bibfield  [0]{\@secondoftwo}\providecommand \translation [1]{[#1]}\providecommand \BibitemOpen [0]{}\providecommand \bibitemStop [0]{}\providecommand \bibitemNoStop [0]{.\EOS\space}\providecommand \EOS [0]{\spacefactor3000\relax}\providecommand \BibitemShut  [1]{\csname bibitem#1\endcsname}\let\auto@bib@innerbib\@empty
\bibitem [{\citenamefont {Michel}(2001)}]{Michel2001}\BibitemOpen
  \bibfield  {author} {\bibinfo {author} {\bibfnamefont {L.}~\bibnamefont
  {Michel}},\ }\bibfield  {title} {\bibinfo {title} {Symmetry, invariants,
  topology. basic tools},\ }\href
  {https://doi.org/10.1016/s0370-1573(00)00088-0} {\bibfield  {journal}
  {\bibinfo  {journal} {Physics Reports}\ }\textbf {\bibinfo {volume} {341}},\
  \bibinfo {pages} {11} (\bibinfo {year} {2001})}\BibitemShut {NoStop}\bibitem [{\citenamefont {Nye}(1985)}]{Nye1985}\BibitemOpen
  \bibfield  {author} {\bibinfo {author} {\bibfnamefont {J.~F.}\ \bibnamefont
  {Nye}},\ }\href@noop {} {\emph {\bibinfo {title} {Physical Properties Of
  Crystals: Their Representation by Tensors and Matrices}}}\ (\bibinfo
  {publisher} {Oxford University Press},\ \bibinfo {year} {1985})\BibitemShut
  {NoStop}\bibitem [{\citenamefont {Malgrange}\ \emph {et~al.}(2014)\citenamefont
  {Malgrange}, \citenamefont {Ricolleau},\ and\ \citenamefont
  {Schlenker}}]{Malgrange2014}\BibitemOpen
  \bibfield  {author} {\bibinfo {author} {\bibfnamefont {C.}~\bibnamefont
  {Malgrange}}, \bibinfo {author} {\bibfnamefont {C.}~\bibnamefont
  {Ricolleau}},\ and\ \bibinfo {author} {\bibfnamefont {M.}~\bibnamefont
  {Schlenker}},\ }\href {https://doi.org/10.1007/978-94-017-8993-6} {\emph
  {\bibinfo {title} {Symmetry and Physical Properties of Crystals}}}\ (\bibinfo
   {publisher} {Springer Netherlands},\ \bibinfo {year} {2014})\BibitemShut
  {NoStop}\bibitem [{\citenamefont {Bertoldi}\ \emph {et~al.}(2017)\citenamefont
  {Bertoldi}, \citenamefont {Vitelli}, \citenamefont {Christensen},\ and\
  \citenamefont {van Hecke}}]{Bertoldi2017}\BibitemOpen
  \bibfield  {author} {\bibinfo {author} {\bibfnamefont {K.}~\bibnamefont
  {Bertoldi}}, \bibinfo {author} {\bibfnamefont {V.}~\bibnamefont {Vitelli}},
  \bibinfo {author} {\bibfnamefont {J.}~\bibnamefont {Christensen}},\ and\
  \bibinfo {author} {\bibfnamefont {M.}~\bibnamefont {van Hecke}},\ }\bibfield
  {title} {\bibinfo {title} {Flexible mechanical metamaterials},\ }\href
  {https://doi.org/10.1038/natrevmats.2017.66} {\bibfield  {journal} {\bibinfo
  {journal} {Nature Reviews Materials}\ }\textbf {\bibinfo {volume} {2}},\
  \bibinfo {pages} {17066} (\bibinfo {year} {2017})}\BibitemShut {NoStop}\bibitem [{\citenamefont {Kadic}\ \emph {et~al.}(2019)\citenamefont {Kadic},
  \citenamefont {Milton}, \citenamefont {van Hecke},\ and\ \citenamefont
  {Wegener}}]{Kadic2019}\BibitemOpen
  \bibfield  {author} {\bibinfo {author} {\bibfnamefont {M.}~\bibnamefont
  {Kadic}}, \bibinfo {author} {\bibfnamefont {G.~W.}\ \bibnamefont {Milton}},
  \bibinfo {author} {\bibfnamefont {M.}~\bibnamefont {van Hecke}},\ and\
  \bibinfo {author} {\bibfnamefont {M.}~\bibnamefont {Wegener}},\ }\bibfield
  {title} {\bibinfo {title} {3d metamaterials},\ }\href
  {https://doi.org/10.1038/s42254-018-0018-y} {\bibfield  {journal} {\bibinfo
  {journal} {Nature Reviews Physics}\ }\textbf {\bibinfo {volume} {1}},\
  \bibinfo {pages} {198} (\bibinfo {year} {2019})}\BibitemShut {NoStop}\bibitem [{\citenamefont {Soukoulis}\ and\ \citenamefont
  {Wegener}(2011)}]{Soukoulis2011}\BibitemOpen
  \bibfield  {author} {\bibinfo {author} {\bibfnamefont {C.~M.}\ \bibnamefont
  {Soukoulis}}\ and\ \bibinfo {author} {\bibfnamefont {M.}~\bibnamefont
  {Wegener}},\ }\bibfield  {title} {\bibinfo {title} {Past achievements and
  future challenges in the development of three-dimensional photonic
  metamaterials},\ }\href {https://doi.org/10.1038/nphoton.2011.154} {\bibfield
   {journal} {\bibinfo  {journal} {Nature Photonics}\ }\textbf {\bibinfo
  {volume} {5}},\ \bibinfo {pages} {523} (\bibinfo {year} {2011})}\BibitemShut
  {NoStop}\bibitem [{\citenamefont {Tolédano}\ and\ \citenamefont
  {Tolédano}(1987)}]{Toledano1987}\BibitemOpen
  \bibfield  {author} {\bibinfo {author} {\bibfnamefont {J.-C.}\ \bibnamefont
  {Tolédano}}\ and\ \bibinfo {author} {\bibfnamefont {P.}~\bibnamefont
  {Tolédano}},\ }\href@noop {} {\emph {\bibinfo {title} {The Landau Theory Of
  Phase Transitions - Application To Structural, Incommensurate, Magnetic, And
  Liquid Crystal Systems}}}\ (\bibinfo  {publisher} {World Scientific
  Publishing Company Incorporated},\ \bibinfo {year} {1987})\BibitemShut
  {NoStop}\bibitem [{\citenamefont {Bradlyn}\ \emph {et~al.}(2017)\citenamefont
  {Bradlyn}, \citenamefont {Elcoro}, \citenamefont {Cano}, \citenamefont
  {Vergniory}, \citenamefont {Wang}, \citenamefont {Felser}, \citenamefont
  {Aroyo},\ and\ \citenamefont {Bernevig}}]{Bradlyn2017}\BibitemOpen
  \bibfield  {author} {\bibinfo {author} {\bibfnamefont {B.}~\bibnamefont
  {Bradlyn}}, \bibinfo {author} {\bibfnamefont {L.}~\bibnamefont {Elcoro}},
  \bibinfo {author} {\bibfnamefont {J.}~\bibnamefont {Cano}}, \bibinfo {author}
  {\bibfnamefont {M.~G.}\ \bibnamefont {Vergniory}}, \bibinfo {author}
  {\bibfnamefont {Z.}~\bibnamefont {Wang}}, \bibinfo {author} {\bibfnamefont
  {C.}~\bibnamefont {Felser}}, \bibinfo {author} {\bibfnamefont {M.~I.}\
  \bibnamefont {Aroyo}},\ and\ \bibinfo {author} {\bibfnamefont {B.~A.}\
  \bibnamefont {Bernevig}},\ }\bibfield  {title} {\bibinfo {title} {Topological
  quantum chemistry},\ }\href {https://doi.org/10.1038/nature23268} {\bibfield
  {journal} {\bibinfo  {journal} {Nature}\ }\textbf {\bibinfo {volume} {547}},\
  \bibinfo {pages} {298} (\bibinfo {year} {2017})}\BibitemShut {NoStop}\bibitem [{\citenamefont {Groot}\ and\ \citenamefont
  {Mazur}(1962)}]{DeGrootMazur}\BibitemOpen
  \bibfield  {author} {\bibinfo {author} {\bibfnamefont {S.~R.~D.}\
  \bibnamefont {Groot}}\ and\ \bibinfo {author} {\bibfnamefont
  {P.}~\bibnamefont {Mazur}},\ }\href@noop {} {\emph {\bibinfo {title}
  {{Non-Equilibrium Thermodynamics}}}}\ (\bibinfo  {publisher} {Dover
  Publications},\ \bibinfo {year} {1962})\BibitemShut {NoStop}\bibitem [{\citenamefont {Lei}\ \emph {et~al.}(2021)\citenamefont {Lei},
  \citenamefont {Zheng}, \citenamefont {Tang}, \citenamefont {Wan},
  \citenamefont {Ni},\ and\ \citenamefont {qiang Ma}}]{Lei2021}\BibitemOpen
  \bibfield  {author} {\bibinfo {author} {\bibfnamefont {Q.-L.}\ \bibnamefont
  {Lei}}, \bibinfo {author} {\bibfnamefont {W.}~\bibnamefont {Zheng}}, \bibinfo
  {author} {\bibfnamefont {F.}~\bibnamefont {Tang}}, \bibinfo {author}
  {\bibfnamefont {X.}~\bibnamefont {Wan}}, \bibinfo {author} {\bibfnamefont
  {R.}~\bibnamefont {Ni}},\ and\ \bibinfo {author} {\bibfnamefont
  {Y.}~\bibnamefont {qiang Ma}},\ }\bibfield  {title} {\bibinfo {title}
  {Self-assembly of isostatic self-dual colloidal crystals},\ }\href
  {https://doi.org/10.1103/physrevlett.127.018001} {\bibfield  {journal}
  {\bibinfo  {journal} {Physical Review Letters}\ }\textbf {\bibinfo {volume}
  {127}},\ \bibinfo {pages} {018001} (\bibinfo {year} {2021})}\BibitemShut
  {NoStop}\bibitem [{\citenamefont {Guarneri}\ \emph {et~al.}(2020)\citenamefont
  {Guarneri}, \citenamefont {Tian},\ and\ \citenamefont {Wang}}]{Guarneri2020}\BibitemOpen
  \bibfield  {author} {\bibinfo {author} {\bibfnamefont {I.}~\bibnamefont
  {Guarneri}}, \bibinfo {author} {\bibfnamefont {C.}~\bibnamefont {Tian}},\
  and\ \bibinfo {author} {\bibfnamefont {J.}~\bibnamefont {Wang}},\ }\bibfield
  {title} {\bibinfo {title} {Self-duality triggered dynamical transition},\
  }\href {https://doi.org/10.1103/physrevb.102.045433} {\bibfield  {journal}
  {\bibinfo  {journal} {Physical Review B}\ }\textbf {\bibinfo {volume}
  {102}},\ \bibinfo {pages} {045433} (\bibinfo {year} {2020})}\BibitemShut
  {NoStop}\bibitem [{\citenamefont {Browne}\ \emph {et~al.}(1984)\citenamefont {Browne},
  \citenamefont {Carini}, \citenamefont {Muttalib},\ and\ \citenamefont
  {Nagel}}]{Browne1984}\BibitemOpen
  \bibfield  {author} {\bibinfo {author} {\bibfnamefont {D.~A.}\ \bibnamefont
  {Browne}}, \bibinfo {author} {\bibfnamefont {J.~P.}\ \bibnamefont {Carini}},
  \bibinfo {author} {\bibfnamefont {K.~A.}\ \bibnamefont {Muttalib}},\ and\
  \bibinfo {author} {\bibfnamefont {S.~R.}\ \bibnamefont {Nagel}},\ }\bibfield
  {title} {\bibinfo {title} {Periodicity of transport coefficients with half
  flux quanta in the aharonov-bohm effect},\ }\href
  {https://doi.org/10.1103/physrevb.30.6798} {\bibfield  {journal} {\bibinfo
  {journal} {Physical Review B}\ }\textbf {\bibinfo {volume} {30}},\ \bibinfo
  {pages} {6798} (\bibinfo {year} {1984})}\BibitemShut {NoStop}\bibitem [{\citenamefont {Hou}(2013)}]{Hou2013}\BibitemOpen
  \bibfield  {author} {\bibinfo {author} {\bibfnamefont {J.-M.}\ \bibnamefont
  {Hou}},\ }\bibfield  {title} {\bibinfo {title} {Hidden-symmetry-protected
  topological semimetals on a square lattice},\ }\href
  {https://doi.org/10.1103/physrevlett.111.130403} {\bibfield  {journal}
  {\bibinfo  {journal} {Physical Review Letters}\ }\textbf {\bibinfo {volume}
  {111}},\ \bibinfo {pages} {130403} (\bibinfo {year} {2013})}\BibitemShut
  {NoStop}\bibitem [{\citenamefont {Fruchart}\ \emph {et~al.}(2020)\citenamefont
  {Fruchart}, \citenamefont {Zhou},\ and\ \citenamefont
  {Vitelli}}]{Fruchart2020}\BibitemOpen
  \bibfield  {author} {\bibinfo {author} {\bibfnamefont {M.}~\bibnamefont
  {Fruchart}}, \bibinfo {author} {\bibfnamefont {Y.}~\bibnamefont {Zhou}},\
  and\ \bibinfo {author} {\bibfnamefont {V.}~\bibnamefont {Vitelli}},\
  }\bibfield  {title} {\bibinfo {title} {Dualities and non-abelian mechanics},\
  }\href {https://doi.org/10.1038/s41586-020-1932-6} {\bibfield  {journal}
  {\bibinfo  {journal} {Nature}\ }\textbf {\bibinfo {volume} {577}},\ \bibinfo
  {pages} {636} (\bibinfo {year} {2020})}\BibitemShut {NoStop}\bibitem [{\citenamefont {Liu}\ and\ \citenamefont
  {Semperlotti}(2021)}]{Liu2021}\BibitemOpen
  \bibfield  {author} {\bibinfo {author} {\bibfnamefont {T.-W.}\ \bibnamefont
  {Liu}}\ and\ \bibinfo {author} {\bibfnamefont {F.}~\bibnamefont
  {Semperlotti}},\ }\bibfield  {title} {\bibinfo {title} {Synthetic kramers
  pair in phononic elastic plates and helical edge states on a dislocation
  interface},\ }\href {https://doi.org/10.1002/adma.202005160} {\bibfield
  {journal} {\bibinfo  {journal} {Advanced Materials}\ }\textbf {\bibinfo
  {volume} {33}},\ \bibinfo {pages} {2005160} (\bibinfo {year}
  {2021})}\BibitemShut {NoStop}\bibitem [{\citenamefont {Danawe}\ \emph {et~al.}(2021)\citenamefont {Danawe},
  \citenamefont {Li}, \citenamefont {Ba'ba'a},\ and\ \citenamefont
  {Tol}}]{Danawe2021}\BibitemOpen
  \bibfield  {author} {\bibinfo {author} {\bibfnamefont {H.}~\bibnamefont
  {Danawe}}, \bibinfo {author} {\bibfnamefont {H.}~\bibnamefont {Li}}, \bibinfo
  {author} {\bibfnamefont {H.~A.}\ \bibnamefont {Ba'ba'a}},\ and\ \bibinfo
  {author} {\bibfnamefont {S.}~\bibnamefont {Tol}},\ }\href@noop {} {\bibinfo
  {title} {Existence of corner modes in elastic twisted kagome lattices}}
  (\bibinfo {year} {2021}),\ \Eprint {https://arxiv.org/abs/2107.07924}
  {arXiv:2107.07924} \BibitemShut {NoStop}\bibitem [{\citenamefont {Gonella}(2020)}]{Gonella2020}\BibitemOpen
  \bibfield  {author} {\bibinfo {author} {\bibfnamefont {S.}~\bibnamefont
  {Gonella}},\ }\bibfield  {title} {\bibinfo {title} {Symmetry of the phononic
  landscape of twisted kagome lattices across the duality boundary},\ }\href
  {https://doi.org/10.1103/physrevb.102.140301} {\bibfield  {journal} {\bibinfo
   {journal} {Physical Review B}\ }\textbf {\bibinfo {volume} {102}},\ \bibinfo
  {pages} {140301} (\bibinfo {year} {2020})}\BibitemShut {NoStop}\bibitem [{\citenamefont {Fruchart}\ and\ \citenamefont
  {Vitelli}(2020)}]{Fruchart2020b}\BibitemOpen
  \bibfield  {author} {\bibinfo {author} {\bibfnamefont {M.}~\bibnamefont
  {Fruchart}}\ and\ \bibinfo {author} {\bibfnamefont {V.}~\bibnamefont
  {Vitelli}},\ }\bibfield  {title} {\bibinfo {title} {Symmetries and dualities
  in the theory of elasticity},\ }\href
  {https://doi.org/10.1103/PhysRevLett.124.248001} {\bibfield  {journal}
  {\bibinfo  {journal} {Physical Review Letters}\ }\textbf {\bibinfo {volume}
  {124}} (\bibinfo {year} {2020})}\BibitemShut {NoStop}\bibitem [{\citenamefont {Matlack}\ \emph {et~al.}(2018)\citenamefont
  {Matlack}, \citenamefont {Serra-Garcia}, \citenamefont {Palermo},
  \citenamefont {Huber},\ and\ \citenamefont {Daraio}}]{Matlack2018}\BibitemOpen
  \bibfield  {author} {\bibinfo {author} {\bibfnamefont {K.~H.}\ \bibnamefont
  {Matlack}}, \bibinfo {author} {\bibfnamefont {M.}~\bibnamefont
  {Serra-Garcia}}, \bibinfo {author} {\bibfnamefont {A.}~\bibnamefont
  {Palermo}}, \bibinfo {author} {\bibfnamefont {S.~D.}\ \bibnamefont {Huber}},\
  and\ \bibinfo {author} {\bibfnamefont {C.}~\bibnamefont {Daraio}},\
  }\bibfield  {title} {\bibinfo {title} {Designing perturbative metamaterials
  from discrete models},\ }\href {https://doi.org/10.1038/s41563-017-0003-3}
  {\bibfield  {journal} {\bibinfo  {journal} {Nature Materials}\ }\textbf
  {\bibinfo {volume} {17}},\ \bibinfo {pages} {323} (\bibinfo {year}
  {2018})}\BibitemShut {NoStop}\bibitem [{\citenamefont {Fruchart}\ and\ \citenamefont
  {Vitelli}(2018)}]{Fruchart2018b}\BibitemOpen
  \bibfield  {author} {\bibinfo {author} {\bibfnamefont {M.}~\bibnamefont
  {Fruchart}}\ and\ \bibinfo {author} {\bibfnamefont {V.}~\bibnamefont
  {Vitelli}},\ }\bibfield  {title} {\bibinfo {title} {The effective way},\
  }\href {https://doi.org/10.1038/s41563-018-0050-4} {\bibfield  {journal}
  {\bibinfo  {journal} {Nature Materials}\ }\textbf {\bibinfo {volume} {17}},\
  \bibinfo {pages} {292} (\bibinfo {year} {2018})}\BibitemShut {NoStop}\bibitem [{\citenamefont {Ozawa}\ \emph {et~al.}(2019)\citenamefont {Ozawa},
  \citenamefont {Price}, \citenamefont {Amo}, \citenamefont {Goldman},
  \citenamefont {Hafezi}, \citenamefont {Lu}, \citenamefont {Rechtsman},
  \citenamefont {Schuster}, \citenamefont {Simon}, \citenamefont {Zilberberg},\
  and\ \citenamefont {Carusotto}}]{Ozawa2019}\BibitemOpen
  \bibfield  {author} {\bibinfo {author} {\bibfnamefont {T.}~\bibnamefont
  {Ozawa}}, \bibinfo {author} {\bibfnamefont {H.~M.}\ \bibnamefont {Price}},
  \bibinfo {author} {\bibfnamefont {A.}~\bibnamefont {Amo}}, \bibinfo {author}
  {\bibfnamefont {N.}~\bibnamefont {Goldman}}, \bibinfo {author} {\bibfnamefont
  {M.}~\bibnamefont {Hafezi}}, \bibinfo {author} {\bibfnamefont
  {L.}~\bibnamefont {Lu}}, \bibinfo {author} {\bibfnamefont {M.~C.}\
  \bibnamefont {Rechtsman}}, \bibinfo {author} {\bibfnamefont {D.}~\bibnamefont
  {Schuster}}, \bibinfo {author} {\bibfnamefont {J.}~\bibnamefont {Simon}},
  \bibinfo {author} {\bibfnamefont {O.}~\bibnamefont {Zilberberg}},\ and\
  \bibinfo {author} {\bibfnamefont {I.}~\bibnamefont {Carusotto}},\ }\bibfield
  {title} {\bibinfo {title} {Topological photonics},\ }\href
  {https://doi.org/10.1103/revmodphys.91.015006} {\bibfield  {journal}
  {\bibinfo  {journal} {Reviews of Modern Physics}\ }\textbf {\bibinfo {volume}
  {91}},\ \bibinfo {pages} {015006} (\bibinfo {year} {2019})}\BibitemShut
  {NoStop}\bibitem [{\citenamefont {Cooper}\ \emph {et~al.}(2019)\citenamefont {Cooper},
  \citenamefont {Dalibard},\ and\ \citenamefont {Spielman}}]{Cooper2019}\BibitemOpen
  \bibfield  {author} {\bibinfo {author} {\bibfnamefont {N.}~\bibnamefont
  {Cooper}}, \bibinfo {author} {\bibfnamefont {J.}~\bibnamefont {Dalibard}},\
  and\ \bibinfo {author} {\bibfnamefont {I.}~\bibnamefont {Spielman}},\
  }\bibfield  {title} {\bibinfo {title} {Topological bands for ultracold
  atoms},\ }\href {https://doi.org/10.1103/revmodphys.91.015005} {\bibfield
  {journal} {\bibinfo  {journal} {Reviews of Modern Physics}\ }\textbf
  {\bibinfo {volume} {91}},\ \bibinfo {pages} {015005} (\bibinfo {year}
  {2019})}\BibitemShut {NoStop}\bibitem [{\citenamefont {Bloch}\ \emph {et~al.}(2008)\citenamefont {Bloch},
  \citenamefont {Dalibard},\ and\ \citenamefont {Zwerger}}]{Bloch2008}\BibitemOpen
  \bibfield  {author} {\bibinfo {author} {\bibfnamefont {I.}~\bibnamefont
  {Bloch}}, \bibinfo {author} {\bibfnamefont {J.}~\bibnamefont {Dalibard}},\
  and\ \bibinfo {author} {\bibfnamefont {W.}~\bibnamefont {Zwerger}},\
  }\bibfield  {title} {\bibinfo {title} {Many-body physics with ultracold
  gases},\ }\href {https://doi.org/10.1103/revmodphys.80.885} {\bibfield
  {journal} {\bibinfo  {journal} {Reviews of Modern Physics}\ }\textbf
  {\bibinfo {volume} {80}},\ \bibinfo {pages} {885} (\bibinfo {year}
  {2008})}\BibitemShut {NoStop}\bibitem [{Note1()}]{Note1}\BibitemOpen
  \bibinfo {note} {In this work, we focus on dualities as defined by
  Eq.~\protect \textup {\hbox {\mathsurround \z@ \protect \normalfont
  (\ignorespaces \ref {duality_def}\unskip \@@italiccorr )}}. This definition
  is inspired by similar (but not identical) kinds of dualities crucial in
  statistical physics~\cite {Savit1980}, mechanics~\cite {Crapo1993,Zhou2019},
  condensed matter physics~\cite {Senthil2004,Zaanen2015,Senthil2019} or
  high-energy physics~\cite {Hull1995,Maldacena1999}.}\BibitemShut {Stop}\bibitem [{Note2()}]{Note2}\BibitemOpen
  \bibinfo {note} {We assume that $f$ is a homeomorphism: a continuous
  bijection with a continuous inverse.}\BibitemShut {Stop}\bibitem [{Note3()}]{Note3}\BibitemOpen
  \bibinfo {note} {We have made the simplifying assumption that the space of
  physical states $\protect \mathcal {H}$ doesn't depend on the parameters.
  This is not necessarily true: in general, $P \times \protect \mathcal {H}$
  may be replaced by vector bundle $\protect \mathcal {E}$ over the parameter
  space~$P$, and~$\protect \mathcal {S}$ may be replaced by the endomorphism
  bundle~$\protect \text {End}(\protect \mathcal {E})$. In this case, dualities
  can be expressed in terms of equivariant vector bundles (see Refs~\cite
  {Segal1968,Merkurjev2005} and references therein for definitions). Starting
  with an abstract duality group $G$, we ask that $\protect \mathcal {E}$ is a
  $G$-equivariant vector bundle. The endomorphism bundle is then also
  $G$-equivariant with the corresponding adjoint action. Let us give a bit more
  detail. The group elements $g \in G$ are an abstract version of the different
  dualities acting on our system; we still have to specify how they act. To do
  so, we consider an action $f : G \times P \to P$ of $G$ on the parameter
  space. Each $g \in G$ gives rise to a function $p \to f_g(p)$ on $P$. This
  specifies how parameters change under the duality~$g$. Then, we define a
  linear action $U : G \times \protect \mathcal {E} \to \protect \mathcal {E}$
  of $G$ on the vector bundle $\protect \mathcal {E}$ of physical states. It
  specifies how states are transformed. We impose that $U_g = U(g,\cdot )$ is a
  linear map from the fiber $\protect \mathcal {E}_p$ over $p$ to the fiber
  $\protect \mathcal {E}_{f_g(p)}$ over~$f_g(p)$. This is the equivariance
  condition, which simply means that a state $\psi $ of the system with
  parameters~$p$ is mapped to a state $U_g \psi $ of the system with
  parameters~$f_g(p)$. We can deduce how operators transform from the way
  states transform. Accordingly, an operator $H$ acting on the states of the
  system with parameters $p$ is mapped to $U_g H U_g^{-1}$, which is an
  operator acting on the states the system with parameters
  $f_g(p)$.}\BibitemShut {Stop}\bibitem [{Note4()}]{Note4}\BibitemOpen
  \bibinfo {note} {We can define a separate order $n'$ such that $U^{n'} =
  \protect \text {Id}$ is the identity operator. In general, $n \not
  =n'$.}\BibitemShut {Stop}\bibitem [{Note5()}]{Note5}\BibitemOpen
  \bibinfo {note} {As a counter-example, consider the function $f(p) = p^2$ on
  the parameter space $P = [-1,1]$, which has infinite order. Similarly,
  consider the diagonal matrix $U = \protect \text {diag}({\protect \rm
  e}^{{\protect \rm i}\alpha }, {\protect \rm e}^{{\protect \rm i}\beta })$.
  Its action by conjugation on a matrix $H$ multiplies the off-diagonal
  elements of $H$ by ${\protect \rm e}^{\pm {\protect \rm i}(\alpha - \beta )}$
  and has infinite order when $\alpha - \beta $ is irrational.}\BibitemShut
  {Stop}\bibitem [{\citenamefont {Bir}\ and\ \citenamefont
  {Pikus}(1975)}]{BirPikus1975}\BibitemOpen
  \bibfield  {author} {\bibinfo {author} {\bibfnamefont {G.~L.}\ \bibnamefont
  {Bir}}\ and\ \bibinfo {author} {\bibfnamefont {G.~E.}\ \bibnamefont
  {Pikus}},\ }\href@noop {} {\emph {\bibinfo {title} {Symmetry and
  Strain-induced Effects in Semiconductors}}}\ (\bibinfo  {publisher} {IPST},\
  \bibinfo {year} {1975})\BibitemShut {NoStop}\bibitem [{\citenamefont {Winkler}(2003)}]{Winkler2003}\BibitemOpen
  \bibfield  {author} {\bibinfo {author} {\bibfnamefont {R.}~\bibnamefont
  {Winkler}},\ }\href {https://doi.org/10.1007/b13586} {\emph {\bibinfo {title}
  {Spin--Orbit Coupling Effects in Two-Dimensional Electron and Hole
  Systems}}}\ (\bibinfo  {publisher} {Springer Berlin Heidelberg},\ \bibinfo
  {year} {2003})\BibitemShut {NoStop}\bibitem [{\citenamefont {Willatzen}\ and\ \citenamefont
  {Voon}(2009)}]{Willatzen2009}\BibitemOpen
  \bibfield  {author} {\bibinfo {author} {\bibfnamefont {M.}~\bibnamefont
  {Willatzen}}\ and\ \bibinfo {author} {\bibfnamefont {L.~C. L.~Y.}\
  \bibnamefont {Voon}},\ }\href {https://doi.org/10.1007/978-3-540-92872-0}
  {\emph {\bibinfo {title} {The k$\cdot$p Method}}}\ (\bibinfo  {publisher}
  {Springer Nature},\ \bibinfo {year} {2009})\BibitemShut {NoStop}\bibitem [{\citenamefont {Pikus}(1961)}]{Pikus1961}\BibitemOpen
  \bibfield  {author} {\bibinfo {author} {\bibfnamefont {G.~E.}\ \bibnamefont
  {Pikus}},\ }\bibfield  {title} {\bibinfo {title} {{A New Method of
  Calculating the Energy Spectrum of Carriers in Semiconductors. II. Account of
  Spin-orbit Interaction}},\ }\href
  {http://www.jetp.ac.ru/cgi-bin/e/index/e/14/5/p1075?a=list} {\bibfield
  {journal} {\bibinfo  {journal} {Soviet Physics JETP}\ }\textbf {\bibinfo
  {volume} {14}} (\bibinfo {year} {1961})}\BibitemShut {NoStop}\bibitem [{\citenamefont {Luttinger}(1956)}]{Luttinger1956}\BibitemOpen
  \bibfield  {author} {\bibinfo {author} {\bibfnamefont {J.~M.}\ \bibnamefont
  {Luttinger}},\ }\bibfield  {title} {\bibinfo {title} {{Quantum Theory of
  Cyclotron Resonance in Semiconductors: General Theory}},\ }\href
  {https://doi.org/10.1103/physrev.102.1030} {\bibfield  {journal} {\bibinfo
  {journal} {Physical Review}\ }\textbf {\bibinfo {volume} {102}},\ \bibinfo
  {pages} {1030} (\bibinfo {year} {1956})}\BibitemShut {NoStop}\bibitem [{\citenamefont {Weyl}(1939)}]{Weyl1939}\BibitemOpen
  \bibfield  {author} {\bibinfo {author} {\bibfnamefont {H.}~\bibnamefont
  {Weyl}},\ }\bibfield  {title} {\bibinfo {title} {Invariants},\ }\href
  {https://doi.org/10.1215/s0012-7094-39-00540-5} {\bibfield  {journal}
  {\bibinfo  {journal} {Duke Mathematical Journal}\ }\textbf {\bibinfo {volume}
  {5}},\ \bibinfo {pages} {489} (\bibinfo {year} {1939})}\BibitemShut {NoStop}\bibitem [{\citenamefont {Sakoda}(2004)}]{Sakoda2004}\BibitemOpen
  \bibfield  {author} {\bibinfo {author} {\bibfnamefont {K.}~\bibnamefont
  {Sakoda}},\ }\href@noop {} {\emph {\bibinfo {title} {{Optical Properties of
  Photonic Crystals}}}},\ \bibinfo {edition} {2nd}\ ed.\ (\bibinfo  {publisher}
  {Springer},\ \bibinfo {year} {2004})\BibitemShut {NoStop}\bibitem [{\citenamefont {Fruchart}\ \emph {et~al.}(2018)\citenamefont
  {Fruchart}, \citenamefont {Jeon}, \citenamefont {Hur}, \citenamefont
  {Cheianov}, \citenamefont {Wiesner},\ and\ \citenamefont
  {Vitelli}}]{Fruchart2018}\BibitemOpen
  \bibfield  {author} {\bibinfo {author} {\bibfnamefont {M.}~\bibnamefont
  {Fruchart}}, \bibinfo {author} {\bibfnamefont {S.-Y.}\ \bibnamefont {Jeon}},
  \bibinfo {author} {\bibfnamefont {K.}~\bibnamefont {Hur}}, \bibinfo {author}
  {\bibfnamefont {V.}~\bibnamefont {Cheianov}}, \bibinfo {author}
  {\bibfnamefont {U.}~\bibnamefont {Wiesner}},\ and\ \bibinfo {author}
  {\bibfnamefont {V.}~\bibnamefont {Vitelli}},\ }\bibfield  {title} {\bibinfo
  {title} {{Soft self-assembly of Weyl materials for light and sound}},\ }\href
  {https://doi.org/10.1073/pnas.1720828115} {\bibfield  {journal} {\bibinfo
  {journal} {Proceedings of the National Academy of Sciences}\ }\textbf
  {\bibinfo {volume} {115}},\ \bibinfo {pages} {E3655} (\bibinfo {year}
  {2018})}\BibitemShut {NoStop}\bibitem [{\citenamefont {Varjas}\ \emph {et~al.}(2018)\citenamefont {Varjas},
  \citenamefont {Rosdahl},\ and\ \citenamefont {Akhmerov}}]{Varjas2018}\BibitemOpen
  \bibfield  {author} {\bibinfo {author} {\bibfnamefont {D.}~\bibnamefont
  {Varjas}}, \bibinfo {author} {\bibfnamefont {T.~O.}\ \bibnamefont
  {Rosdahl}},\ and\ \bibinfo {author} {\bibfnamefont {A.~R.}\ \bibnamefont
  {Akhmerov}},\ }\bibfield  {title} {\bibinfo {title} {Qsymm: algorithmic
  symmetry finding and symmetric hamiltonian generation},\ }\href
  {https://doi.org/10.1088/1367-2630/aadf67} {\bibfield  {journal} {\bibinfo
  {journal} {New Journal of Physics}\ }\textbf {\bibinfo {volume} {20}},\
  \bibinfo {pages} {093026} (\bibinfo {year} {2018})}\BibitemShut {NoStop}\bibitem [{\citenamefont {Gresch}(2018)}]{Gresch2018}\BibitemOpen
  \bibfield  {author} {\bibinfo {author} {\bibfnamefont {D.}~\bibnamefont
  {Gresch}},\ }\emph {\bibinfo {title} {Identifying Topological Semimetals}},\
  \href {https://doi.org/10.3929/ETHZ-B-000308602} {Ph.D. thesis} (\bibinfo
  {year} {2018})\BibitemShut {NoStop}\bibitem [{\citenamefont {Chertkov}\ \emph {et~al.}(2020)\citenamefont
  {Chertkov}, \citenamefont {Villalonga},\ and\ \citenamefont
  {Clark}}]{Chertkov2020}\BibitemOpen
  \bibfield  {author} {\bibinfo {author} {\bibfnamefont {E.}~\bibnamefont
  {Chertkov}}, \bibinfo {author} {\bibfnamefont {B.}~\bibnamefont
  {Villalonga}},\ and\ \bibinfo {author} {\bibfnamefont {B.~K.}\ \bibnamefont
  {Clark}},\ }\bibfield  {title} {\bibinfo {title} {Engineering topological
  models with a general-purpose symmetry-to-hamiltonian approach},\ }\href
  {https://doi.org/10.1103/physrevresearch.2.023348} {\bibfield  {journal}
  {\bibinfo  {journal} {Physical Review Research}\ }\textbf {\bibinfo {volume}
  {2}},\ \bibinfo {pages} {023348} (\bibinfo {year} {2020})}\BibitemShut
  {NoStop}\bibitem [{Note6()}]{Note6}\BibitemOpen
  \bibinfo {note} {Namely, we ask that there is another homeomorphisms $g$ such
  that $g \circ f \circ g^{-1}$ is an isometry, see Ref.~\cite
  {Kuznetsov2004}.}\BibitemShut {Stop}\bibitem [{Note7()}]{Note7}\BibitemOpen
  \bibinfo {note} {Because we have assumed the duality to have finite order, we
  consider only periodic homeomorphisms (homeomorphisms $f$ such that $f^{\circ
  m} = \protect \text {id}$ for a finite integer $m$). (The case of
  infinite-order dualities is outside of the scope of this work.) This
  hypothesis, along with that $f$ is an isometry, gives relatively strong
  constraints on the duality map. If we are only interested about local (i.e.,
  not global) properties in parameter space, this implies that $f$ is linear.
  This can be seen from a series expansion near a self-dual point $p_0$ as
  follows. Write $f(p) = f(p_0 + \delta p) = p_0 + F \protect \tmspace
  +\thinmuskip {.1667em} \delta p + \protect \mathcal {O}(\delta p^2)$ where
  $\delta p = p - p_0$. The matrix $F$ satisfies $F^m = \protect \text {Id}$
  (because $f^{\circ m} = \protect \text {id}$), and is therefore orthogonal.
  The case of global properties is considerably more complicated; however,
  several results exist in simple cases. For instance, all the periodic
  homeomorphism of the real line $\protect \mathbb {R}$ or on the closed
  interval $[-1, 1]$ are either the identity map $\protect \text {id}$ or
  topologically conjugate to the reflection map $x \DOTSB \mapstochar
  \rightarrow -x$. We direct the reader to Ref.~\cite {Constantin2003} and
  references therein for more details.}\BibitemShut {Stop}\bibitem [{Note8()}]{Note8}\BibitemOpen
  \bibinfo {note} {This is a consequence of the Mazur-Ulam theorem and its
  generalizations, when applicable, as we have assumed that the duality map is
  an isometry. We direct the reader to Refs.~\cite
  {Mazur1932,Vaisala2003,Molnar2015} and references therein for more
  details.}\BibitemShut {Stop}\bibitem [{Note9()}]{Note9}\BibitemOpen
  \bibinfo {note} {We get $\protect \mathaccentV {hat}05E{U} \protect
  \mathaccentV {hat}05E{H}_{\gamma }(F p) \protect \mathaccentV
  {hat}05E{U}^{-1} = \delta _{\mu \nu } \protect \tmspace +\thinmuskip
  {.1667em} \protect \overline {\rho (\gamma )}_{\mu \mu '} \protect \tmspace
  +\thinmuskip {.1667em} \rho (\gamma )_{\nu \nu '} \protect \tmspace
  +\thinmuskip {.1667em} a_{\gamma }^{\mu '}(p) \protect \tmspace +\thinmuskip
  {.1667em} \protect \mathaccentV {hat}05E{H}_{\gamma }^{\nu '}$ and as $\rho
  (\gamma )$ is unitary, we have $\delta _{\mu \nu } \protect \overline {\rho
  (\gamma )_{\mu \mu '}} \rho (\gamma )_{\nu \nu '} = [\rho (\gamma )^\dagger
  \rho (\gamma )]_{\mu ' \nu '} = \delta _{\mu ' \nu '}$ which gives the
  result.}\BibitemShut {Stop}\bibitem [{Note10()}]{Note10}\BibitemOpen
  \bibinfo {note} {To make contact with equivalent notations, note that the
  translation operator can be written $\protect \mathaccentV {hat}05E{T}(\gamma
  ) = \DOTSB \sum@ \slimits@ _{x \in \protect \mathcal {C}} \mathinner
  {|{x+\gamma }\delimiter "526930B }\protect \tmspace -\thinmuskip
  {.1667em}\mathinner {\delimiter "426830A {x}|}$ or equivalently $\protect
  \mathaccentV {hat}05E{T}(\gamma ) = \DOTSB \sum@ \slimits@ _{x \in \protect
  \mathcal {C}} \protect \mathaccentV {hat}05E{c}^\dagger _{x+\gamma }\protect
  \tmspace +\thinmuskip {.1667em} \protect \mathaccentV {hat}05E{c}_x$ where
  $\mathinner {|{x}\delimiter "526930B }$ is a state fully localized at the
  point $x$ of the crystal, and $\protect \mathaccentV {hat}05E{c}_x^{(\dagger
  )}$ the corresponding annihilation (creation) operator. Given a basis
  $\mathinner {|{e_i}\delimiter "526930B }$ of the Hilbert space of the degrees
  of freedom in the unit cell, we can further decompose $h(\gamma , p) = \DOTSB
  \sum@ \slimits@ _{} h_{i j}(\gamma , p) \mathinner {|{e_i}\delimiter "526930B
  }\protect \tmspace -\thinmuskip {.1667em}\mathinner {\delimiter "426830A
  {e_j}|}$ in which the matrix elements $h_{i j}(\gamma , p)$ are now scalars.
  The Hamiltonians $\protect \mathaccentV {hat}05E{H}(p)$ act on the Hilbert
  space spanned by states of the form $\mathinner {|{x}\delimiter "526930B }
  \otimes \mathinner {|{e_i}\delimiter "526930B }$.}\BibitemShut {Stop}\bibitem [{Note11()}]{Note11}\BibitemOpen
  \bibinfo {note} {We can also write more explicitly \begin {equation} \protect
  \mathaccentV {hat}05E{U} = \DOTSB \sum@ \slimits@ _{{\protect \substack
  {\gamma ,\gamma ' \in \Gamma \\ \xi ,\xi ' \in \protect \mathscr {F}}}}
  u_{\xi ',\xi }(\gamma ', \gamma ) \mathinner {|{\gamma ', \xi '}\delimiter
  "526930B } \mathinner {\delimiter "426830A {\gamma , \xi }|} \end {equation}
  in which $\mathinner {|{\gamma , \xi }\delimiter "526930B }$ are a basis of
  $\ell ^2(\protect \mathcal {C})$ of functions $x \DOTSB \mapstochar
  \rightarrow \delta (x - (\gamma + \xi ))$ (for $\gamma \in \Gamma $ and $\xi
  \in \protect \mathscr {F}$, where $\protect \mathscr {F}$ is a fundamental
  domain) having value one at $\gamma + \xi \in \protect \mathcal {C}$ and zero
  elsewhere.}\BibitemShut {Stop}\bibitem [{Note12()}]{Note12}\BibitemOpen
  \bibinfo {note} {One can further convert Eq.~\protect \textup {\hbox
  {\mathsurround \z@ \protect \normalfont (\ignorespaces \ref
  {linear_eq_duality_AB}\unskip \@@italiccorr )}} into the homogeneous linear
  system $[\protect \mathcal {A}, \protect \mathcal {B}] \protect \tmspace
  +\thinmuskip {.1667em} [\protect \mathaccentV {vec}17E{h}(p), \protect
  \mathaccentV {vec}17E{h}(f(p))]^T = 0$. The vectors and matrices are finite,
  provided that we restrict the model to have finite range connections
  (corresponding to the restriction to $\protect \ensuremath {\Gamma
  _{0}}$).}\BibitemShut {Stop}\bibitem [{\citenamefont {Geddes}\ \emph {et~al.}(1992)\citenamefont {Geddes},
  \citenamefont {Czapor},\ and\ \citenamefont {Labahn}}]{Geddes1992}\BibitemOpen
  \bibfield  {author} {\bibinfo {author} {\bibfnamefont {K.~O.}\ \bibnamefont
  {Geddes}}, \bibinfo {author} {\bibfnamefont {S.~R.}\ \bibnamefont {Czapor}},\
  and\ \bibinfo {author} {\bibfnamefont {G.}~\bibnamefont {Labahn}},\ }\href
  {https://doi.org/10.1007/b102438} {\emph {\bibinfo {title} {Algorithms for
  Computer Algebra}}}\ (\bibinfo  {publisher} {Springer {US}},\ \bibinfo {year}
  {1992})\BibitemShut {NoStop}\bibitem [{\citenamefont {Press}(2007)}]{Press2007}\BibitemOpen
  \bibinfo {editor} {\bibfnamefont {W.~H.}\ \bibnamefont {Press}},\ ed.,\
  \href@noop {} {\emph {\bibinfo {title} {Numerical recipes: the art of
  scientific computing}}},\ \bibinfo {edition} {3rd}\ ed.\ (\bibinfo
  {publisher} {Cambridge University Press},\ \bibinfo {address} {Cambridge, UK
  ; New York},\ \bibinfo {year} {2007})\BibitemShut {NoStop}\bibitem [{Note13()}]{Note13}\BibitemOpen
  \bibinfo {note} {It is still possible to directly solve the duality equation
  when $f$ is non-linear, but we don't have a systematic way of writing an
  explicit parametrization.}\BibitemShut {Stop}\bibitem [{Note14()}]{Note14}\BibitemOpen
  \bibinfo {note} {We have shortened to $U(\protect \mathcal {O} k)$ the
  momentum-dependent duality operator $U(k, \protect \mathcal {O} k)$. This is
  a choice: we could have called the same quantity $U(k)$.}\BibitemShut {Stop}\bibitem [{Note15()}]{Note15}\BibitemOpen
  \bibinfo {note} {This is contingent on the existence of free will. We refer
  to the articles~\cite {Conway2006,Chiang2005,Aaronson2016} for
  discussions.}\BibitemShut {Stop}\bibitem [{Note16()}]{Note16}\BibitemOpen
  \bibinfo {note} {It is even dubious that the notation $D_1 + D_2$ has any
  meaning, because $D_1$ and $D_2$ act on different spaces (that are not
  canonically isomorphic).}\BibitemShut {Stop}\bibitem [{Note17()}]{Note17}\BibitemOpen
  \bibinfo {note} {In practice, it is convenient to focus on the cases in which
  there is a finite number of translation operators in Eq.~\protect \textup
  {\hbox {\mathsurround \z@ \protect \normalfont (\ignorespaces \ref
  {function_F_def}\unskip \@@italiccorr )}} (see the discussion after
  Eq.~\protect \textup {\hbox {\mathsurround \z@ \protect \normalfont
  (\ignorespaces \ref {generic_hamiltonian}\unskip \@@italiccorr )}}), so that
  \begin {equation} \protect \mathcal {F}(s) = \DOTSB \sum@ \slimits@ _{\gamma
  \in \Gamma } \protect \mathcal {F}(s; \gamma ) T(\gamma ) \end {equation} is
  a finite sum, in which $\protect \mathcal {F}(s; \gamma )$ are
  finite-dimensional matrices.}\BibitemShut {Stop}\bibitem [{\citenamefont {Allgower}\ and\ \citenamefont
  {Georg}(2003)}]{Allgower2003}\BibitemOpen
  \bibfield  {author} {\bibinfo {author} {\bibfnamefont {E.}~\bibnamefont
  {Allgower}}\ and\ \bibinfo {author} {\bibfnamefont {K.}~\bibnamefont
  {Georg}},\ }\href@noop {} {\emph {\bibinfo {title} {Introduction to Numerical
  Continuation Methods}}},\ Classics in Applied Mathematics\ (\bibinfo
  {publisher} {Society for Industrial and Applied Mathematics},\ \bibinfo
  {year} {2003})\BibitemShut {NoStop}\bibitem [{\citenamefont {Kuznetsov}(2004)}]{Kuznetsov2004}\BibitemOpen
  \bibfield  {author} {\bibinfo {author} {\bibfnamefont {Y.~A.}\ \bibnamefont
  {Kuznetsov}},\ }\href {https://doi.org/10.1007/978-1-4757-3978-7} {\emph
  {\bibinfo {title} {Elements of Applied Bifurcation Theory}}}\ (\bibinfo
  {publisher} {Springer New York},\ \bibinfo {year} {2004})\BibitemShut
  {NoStop}\bibitem [{\citenamefont {Li}\ \emph {et~al.}(2019)\citenamefont {Li},
  \citenamefont {Peng}, \citenamefont {Han}, \citenamefont {Miri},
  \citenamefont {Li}, \citenamefont {Xiao}, \citenamefont {Zhu}, \citenamefont
  {Zhao}, \citenamefont {Al{\`u}}, \citenamefont {Fan},\ and\ \citenamefont
  {Qiu}}]{Li2019}\BibitemOpen
  \bibfield  {author} {\bibinfo {author} {\bibfnamefont {Y.}~\bibnamefont
  {Li}}, \bibinfo {author} {\bibfnamefont {Y.-G.}\ \bibnamefont {Peng}},
  \bibinfo {author} {\bibfnamefont {L.}~\bibnamefont {Han}}, \bibinfo {author}
  {\bibfnamefont {M.-A.}\ \bibnamefont {Miri}}, \bibinfo {author}
  {\bibfnamefont {W.}~\bibnamefont {Li}}, \bibinfo {author} {\bibfnamefont
  {M.}~\bibnamefont {Xiao}}, \bibinfo {author} {\bibfnamefont {X.-F.}\
  \bibnamefont {Zhu}}, \bibinfo {author} {\bibfnamefont {J.}~\bibnamefont
  {Zhao}}, \bibinfo {author} {\bibfnamefont {A.}~\bibnamefont {Al{\`u}}},
  \bibinfo {author} {\bibfnamefont {S.}~\bibnamefont {Fan}},\ and\ \bibinfo
  {author} {\bibfnamefont {C.-W.}\ \bibnamefont {Qiu}},\ }\bibfield  {title}
  {\bibinfo {title} {Anti-parity-time symmetry in diffusive systems},\ }\href
  {https://doi.org/10.1126/science.aaw6259} {\bibfield  {journal} {\bibinfo
  {journal} {Science}\ }\textbf {\bibinfo {volume} {364}},\ \bibinfo {pages}
  {170} (\bibinfo {year} {2019})}\BibitemShut {NoStop}\bibitem [{\citenamefont {Ma}\ \emph {et~al.}(2019)\citenamefont {Ma},
  \citenamefont {Xiao},\ and\ \citenamefont {Chan}}]{Ma2019}\BibitemOpen
  \bibfield  {author} {\bibinfo {author} {\bibfnamefont {G.}~\bibnamefont
  {Ma}}, \bibinfo {author} {\bibfnamefont {M.}~\bibnamefont {Xiao}},\ and\
  \bibinfo {author} {\bibfnamefont {C.~T.}\ \bibnamefont {Chan}},\ }\bibfield
  {title} {\bibinfo {title} {Topological phases in acoustic and mechanical
  systems},\ }\href {https://doi.org/10.1038/s42254-019-0030-x} {\bibfield
  {journal} {\bibinfo  {journal} {Nature Reviews Physics}\ }\textbf {\bibinfo
  {volume} {1}},\ \bibinfo {pages} {281} (\bibinfo {year} {2019})}\BibitemShut
  {NoStop}\bibitem [{\citenamefont {Nassar}\ \emph {et~al.}(2020)\citenamefont {Nassar},
  \citenamefont {Yousefzadeh}, \citenamefont {Fleury}, \citenamefont {Ruzzene},
  \citenamefont {Al{\`{u}}}, \citenamefont {Daraio}, \citenamefont {Norris},
  \citenamefont {Huang},\ and\ \citenamefont {Haberman}}]{Nassar2020}\BibitemOpen
  \bibfield  {author} {\bibinfo {author} {\bibfnamefont {H.}~\bibnamefont
  {Nassar}}, \bibinfo {author} {\bibfnamefont {B.}~\bibnamefont {Yousefzadeh}},
  \bibinfo {author} {\bibfnamefont {R.}~\bibnamefont {Fleury}}, \bibinfo
  {author} {\bibfnamefont {M.}~\bibnamefont {Ruzzene}}, \bibinfo {author}
  {\bibfnamefont {A.}~\bibnamefont {Al{\`{u}}}}, \bibinfo {author}
  {\bibfnamefont {C.}~\bibnamefont {Daraio}}, \bibinfo {author} {\bibfnamefont
  {A.~N.}\ \bibnamefont {Norris}}, \bibinfo {author} {\bibfnamefont
  {G.}~\bibnamefont {Huang}},\ and\ \bibinfo {author} {\bibfnamefont {M.~R.}\
  \bibnamefont {Haberman}},\ }\bibfield  {title} {\bibinfo {title}
  {Nonreciprocity in acoustic and elastic materials},\ }\href
  {https://doi.org/10.1038/s41578-020-0206-0} {\bibfield  {journal} {\bibinfo
  {journal} {Nature Reviews Materials}\ }\textbf {\bibinfo {volume} {5}},\
  \bibinfo {pages} {667} (\bibinfo {year} {2020})}\BibitemShut {NoStop}\bibitem [{\citenamefont {Blount}(1962)}]{Blount1962}\BibitemOpen
  \bibfield  {author} {\bibinfo {author} {\bibfnamefont {E.}~\bibnamefont
  {Blount}},\ }\bibfield  {title} {\bibinfo {title} {Formalisms of band
  theory},\ }in\ \href {https://doi.org/10.1016/s0081-1947(08)60459-2} {\emph
  {\bibinfo {booktitle} {Solid State Physics}}}\ (\bibinfo  {publisher}
  {Elsevier},\ \bibinfo {year} {1962})\ pp.\ \bibinfo {pages}
  {305--373}\BibitemShut {NoStop}\bibitem [{\citenamefont {Zak}(1967)}]{Zak1967}\BibitemOpen
  \bibfield  {author} {\bibinfo {author} {\bibfnamefont {J.}~\bibnamefont
  {Zak}},\ }\bibfield  {title} {\bibinfo {title} {Finite translations in
  solid-state physics},\ }\href {https://doi.org/10.1103/physrevlett.19.1385}
  {\bibfield  {journal} {\bibinfo  {journal} {Physical Review Letters}\
  }\textbf {\bibinfo {volume} {19}},\ \bibinfo {pages} {1385} (\bibinfo {year}
  {1967})}\BibitemShut {NoStop}\bibitem [{\citenamefont {Zak}(1989)}]{Zak1989}\BibitemOpen
  \bibfield  {author} {\bibinfo {author} {\bibfnamefont {J.}~\bibnamefont
  {Zak}},\ }\bibfield  {title} {\bibinfo {title} {Berry's phase for energy
  bands in solids},\ }\href {https://doi.org/10.1103/physrevlett.62.2747}
  {\bibfield  {journal} {\bibinfo  {journal} {Physical Review Letters}\
  }\textbf {\bibinfo {volume} {62}},\ \bibinfo {pages} {2747} (\bibinfo {year}
  {1989})}\BibitemShut {NoStop}\bibitem [{\citenamefont {Panati}\ \emph {et~al.}(2003)\citenamefont {Panati},
  \citenamefont {Spohn},\ and\ \citenamefont {Teufel}}]{Panati2003}\BibitemOpen
  \bibfield  {author} {\bibinfo {author} {\bibfnamefont {G.}~\bibnamefont
  {Panati}}, \bibinfo {author} {\bibfnamefont {H.}~\bibnamefont {Spohn}},\ and\
  \bibinfo {author} {\bibfnamefont {S.}~\bibnamefont {Teufel}},\ }\bibfield
  {title} {\bibinfo {title} {Effective dynamics for bloch electrons: Peierls
  substitution and beyond},\ }\href {https://doi.org/10.1007/s00220-003-0950-1}
  {\bibfield  {journal} {\bibinfo  {journal} {Communications in Mathematical
  Physics}\ }\textbf {\bibinfo {volume} {242}},\ \bibinfo {pages} {547}
  (\bibinfo {year} {2003})}\BibitemShut {NoStop}\bibitem [{\citenamefont {Bena}\ and\ \citenamefont
  {Montambaux}(2009)}]{Bena2009}\BibitemOpen
  \bibfield  {author} {\bibinfo {author} {\bibfnamefont {C.}~\bibnamefont
  {Bena}}\ and\ \bibinfo {author} {\bibfnamefont {G.}~\bibnamefont
  {Montambaux}},\ }\bibfield  {title} {\bibinfo {title} {Remarks on the
  tight-binding model of graphene},\ }\href
  {https://doi.org/10.1088/1367-2630/11/9/095003} {\bibfield  {journal}
  {\bibinfo  {journal} {New Journal of Physics}\ }\textbf {\bibinfo {volume}
  {11}},\ \bibinfo {pages} {095003} (\bibinfo {year} {2009})}\BibitemShut
  {NoStop}\bibitem [{\citenamefont {Fruchart}\ \emph {et~al.}(2014)\citenamefont
  {Fruchart}, \citenamefont {Carpentier},\ and\ \citenamefont
  {Gaw{\k{e}}dzki}}]{Fruchart2014}\BibitemOpen
  \bibfield  {author} {\bibinfo {author} {\bibfnamefont {M.}~\bibnamefont
  {Fruchart}}, \bibinfo {author} {\bibfnamefont {D.}~\bibnamefont
  {Carpentier}},\ and\ \bibinfo {author} {\bibfnamefont {K.}~\bibnamefont
  {Gaw{\k{e}}dzki}},\ }\bibfield  {title} {\bibinfo {title} {Parallel transport
  and band theory in crystals},\ }\href
  {https://doi.org/10.1209/0295-5075/106/60002} {\bibfield  {journal} {\bibinfo
   {journal} {{EPL} (Europhysics Letters)}\ }\textbf {\bibinfo {volume}
  {106}},\ \bibinfo {pages} {60002} (\bibinfo {year} {2014})}\BibitemShut
  {NoStop}\bibitem [{\citenamefont {Dobard{\v{z}}i{\'{c}}}\ \emph
  {et~al.}(2014)\citenamefont {Dobard{\v{z}}i{\'{c}}}, \citenamefont
  {Dimitrijevi{\'{c}}},\ and\ \citenamefont {Milovanovi{\'{c}}}}]{Dobardi2014}\BibitemOpen
  \bibfield  {author} {\bibinfo {author} {\bibfnamefont {E.}~\bibnamefont
  {Dobard{\v{z}}i{\'{c}}}}, \bibinfo {author} {\bibfnamefont {M.}~\bibnamefont
  {Dimitrijevi{\'{c}}}},\ and\ \bibinfo {author} {\bibfnamefont {M.~V.}\
  \bibnamefont {Milovanovi{\'{c}}}},\ }\bibfield  {title} {\bibinfo {title}
  {Effective description of chern insulators},\ }\href
  {https://doi.org/10.1103/physrevb.89.235424} {\bibfield  {journal} {\bibinfo
  {journal} {Physical Review B}\ }\textbf {\bibinfo {volume} {89}},\ \bibinfo
  {pages} {235424} (\bibinfo {year} {2014})}\BibitemShut {NoStop}\bibitem [{\citenamefont {Dobard{\v{z}}i{\'{c}}}\ \emph
  {et~al.}(2015)\citenamefont {Dobard{\v{z}}i{\'{c}}}, \citenamefont
  {Dimitrijevi{\'{c}}},\ and\ \citenamefont {Milovanovi{\'{c}}}}]{Dobardi2015}\BibitemOpen
  \bibfield  {author} {\bibinfo {author} {\bibfnamefont {E.}~\bibnamefont
  {Dobard{\v{z}}i{\'{c}}}}, \bibinfo {author} {\bibfnamefont {M.}~\bibnamefont
  {Dimitrijevi{\'{c}}}},\ and\ \bibinfo {author} {\bibfnamefont {M.~V.}\
  \bibnamefont {Milovanovi{\'{c}}}},\ }\bibfield  {title} {\bibinfo {title}
  {Generalized bloch theorem and topological characterization},\ }\href
  {https://doi.org/10.1103/physrevb.91.125424} {\bibfield  {journal} {\bibinfo
  {journal} {Physical Review B}\ }\textbf {\bibinfo {volume} {91}},\ \bibinfo
  {pages} {125424} (\bibinfo {year} {2015})}\BibitemShut {NoStop}\bibitem [{\citenamefont {Lim}\ \emph {et~al.}(2015)\citenamefont {Lim},
  \citenamefont {Fuchs},\ and\ \citenamefont {Montambaux}}]{Lim2015}\BibitemOpen
  \bibfield  {author} {\bibinfo {author} {\bibfnamefont {L.-K.}\ \bibnamefont
  {Lim}}, \bibinfo {author} {\bibfnamefont {J.-N.}\ \bibnamefont {Fuchs}},\
  and\ \bibinfo {author} {\bibfnamefont {G.}~\bibnamefont {Montambaux}},\
  }\bibfield  {title} {\bibinfo {title} {Geometry of bloch states probed by
  stückelberg interferometry},\ }\href
  {https://doi.org/10.1103/physreva.92.063627} {\bibfield  {journal} {\bibinfo
  {journal} {Physical Review A}\ }\textbf {\bibinfo {volume} {92}},\ \bibinfo
  {pages} {063627} (\bibinfo {year} {2015})}\BibitemShut {NoStop}\bibitem [{\citenamefont {Yusufaly}\ \emph {et~al.}(2018)\citenamefont
  {Yusufaly}, \citenamefont {Vanderbilt},\ and\ \citenamefont
  {Coh}}]{PythTB2018}\BibitemOpen
  \bibfield  {author} {\bibinfo {author} {\bibfnamefont {T.}~\bibnamefont
  {Yusufaly}}, \bibinfo {author} {\bibfnamefont {D.}~\bibnamefont
  {Vanderbilt}},\ and\ \bibinfo {author} {\bibfnamefont {S.}~\bibnamefont
  {Coh}},\ }\bibfield  {title} {\bibinfo {title} {{Tight-Binding Formalism in
  the Context of the PythTB Package}}} (\bibinfo {year} {2018})\BibitemShut
  {NoStop}\bibitem [{Note18()}]{Note18}\BibitemOpen
  \bibinfo {note} {The order of the indices might seem strange. This is because
  we have already decided that $\mathinner {\delimiter "426830A {H}\delimiter
  "526930B }_{i j} = \mathinner {\delimiter "426830A {e_i, \protect
  \mathaccentV {hat}05E{H} e_j}\delimiter "526930B }$ and hence that $\protect
  \mathaccentV {hat}05E{H} = \DOTSB \sum@ \slimits@ \mathinner
  {|{e_i}\delimiter "526930B } \protect \tmspace +\thinmuskip {.1667em} H_{i j}
  \mathinner {\delimiter "426830A {e_j}|}$. As a consequence, we must have
  $\protect \mathaccentV {hat}05E{H} \mathinner {|{e_j}\delimiter "526930B } =
  H_{j i} \mathinner {|{e_i}\delimiter "526930B }$. The impossibility of
  ordering indices in a natural way in both expressions at the same time is
  clear evidence of the mercifulness of Yog-Sothoth, who hides from our sight
  the things-that-should-not-be-seen and from our understanding the
  things-that-should-not-be-understood.}\BibitemShut {Stop}\bibitem [{\citenamefont {Varjas}\ \emph {et~al.}(2015)\citenamefont {Varjas},
  \citenamefont {de~Juan},\ and\ \citenamefont {Lu}}]{Varjas2015}\BibitemOpen
  \bibfield  {author} {\bibinfo {author} {\bibfnamefont {D.}~\bibnamefont
  {Varjas}}, \bibinfo {author} {\bibfnamefont {F.}~\bibnamefont {de~Juan}},\
  and\ \bibinfo {author} {\bibfnamefont {Y.-M.}\ \bibnamefont {Lu}},\
  }\bibfield  {title} {\bibinfo {title} {Bulk invariants and topological
  response in insulators and superconductors with nonsymmorphic symmetries},\
  }\href {https://doi.org/10.1103/physrevb.92.195116} {\bibfield  {journal}
  {\bibinfo  {journal} {Physical Review B}\ }\textbf {\bibinfo {volume} {92}},\
  \bibinfo {pages} {195116} (\bibinfo {year} {2015})}\BibitemShut {NoStop}\bibitem [{\citenamefont {Bradley}\ and\ \citenamefont
  {Cracknell}(2010)}]{BradleyCracknell}\BibitemOpen
  \bibfield  {author} {\bibinfo {author} {\bibfnamefont {C.}~\bibnamefont
  {Bradley}}\ and\ \bibinfo {author} {\bibfnamefont {A.}~\bibnamefont
  {Cracknell}},\ }\href@noop {} {\emph {\bibinfo {title} {The Mathematical
  Theory of Symmetry in Solids: Representation Theory for Point Groups and
  Space Groups}}}\ (\bibinfo  {publisher} {Oxford University Press},\ \bibinfo
  {year} {2010})\BibitemShut {NoStop}\bibitem [{\citenamefont {Aroyo}(2016)}]{ITA}\BibitemOpen
  \bibinfo {editor} {\bibfnamefont {M.~I.}\ \bibnamefont {Aroyo}},\ ed.,\ \href
  {https://doi.org/10.1107/97809553602060000114} {\emph {\bibinfo {title}
  {{International Tables for Crystallography, Volume A: Space-group
  symmetry}}}}\ (\bibinfo  {publisher} {International Union of
  Crystallography},\ \bibinfo {year} {2016})\BibitemShut {NoStop}\bibitem [{\citenamefont {El-Batanouny}\ and\ \citenamefont
  {Wooten}(2008)}]{el2008symmetry}\BibitemOpen
  \bibfield  {author} {\bibinfo {author} {\bibfnamefont {M.}~\bibnamefont
  {El-Batanouny}}\ and\ \bibinfo {author} {\bibfnamefont {F.}~\bibnamefont
  {Wooten}},\ }\href {https://books.google.com/books?id=QU-1ngEACAAJ} {\emph
  {\bibinfo {title} {Symmetry and Condensed Matter Physics: A Computational
  Approach}}}\ (\bibinfo  {publisher} {Cambridge University Press},\ \bibinfo
  {year} {2008})\BibitemShut {NoStop}\bibitem [{\citenamefont {Opechowski}(1986)}]{Opechowski1986}\BibitemOpen
  \bibfield  {author} {\bibinfo {author} {\bibfnamefont {W.}~\bibnamefont
  {Opechowski}},\ }\href@noop {} {\emph {\bibinfo {title} {Crystallographic And
  Metacrystallographic Groups}}}\ (\bibinfo  {publisher} {North-Holland Physics
  Publishing},\ \bibinfo {year} {1986})\BibitemShut {NoStop}\bibitem [{Note19()}]{Note19}\BibitemOpen
  \bibinfo {note} {This requires the following property: point group symmetries
  preserve the Bravais lattice. This is because the point group of a crystal is
  a subgroup of the point group of the Bravais lattice (called the holohedry
  group). See for instance \cite [\protect \S ~1.5 p.~40]{BradleyCracknell} or
  \cite [\protect \S ~10.3.6 p.~291]{el2008symmetry}. This shows $R \Gamma
  \subset \Gamma $. As $b_{j}(g)$ is a Bravais lattice vector by construction,
  $\gamma '$ is indeed a Bravais lattice vector.}\BibitemShut {Stop}\bibitem [{\citenamefont {Wigner}(1960)}]{Wigner1960}\BibitemOpen
  \bibfield  {author} {\bibinfo {author} {\bibfnamefont {E.~P.}\ \bibnamefont
  {Wigner}},\ }\bibfield  {title} {\bibinfo {title} {Normal form of antiunitary
  operators},\ }\href {https://doi.org/10.1063/1.1703672} {\bibfield  {journal}
  {\bibinfo  {journal} {Journal of Mathematical Physics}\ }\textbf {\bibinfo
  {volume} {1}},\ \bibinfo {pages} {409} (\bibinfo {year} {1960})}\BibitemShut
  {NoStop}\bibitem [{\citenamefont {Weigert}(2003)}]{Weigert2003}\BibitemOpen
  \bibfield  {author} {\bibinfo {author} {\bibfnamefont {S.}~\bibnamefont
  {Weigert}},\ }\bibfield  {title} {\bibinfo {title} {Pt-symmetry and its
  spontaneous breakdown explained by anti-linearity},\ }\href
  {https://doi.org/10.1088/1464-4266/5/3/380} {\bibfield  {journal} {\bibinfo
  {journal} {Journal of Optics B: Quantum and Semiclassical Optics}\ }\textbf
  {\bibinfo {volume} {5}},\ \bibinfo {pages} {S416} (\bibinfo {year}
  {2003})}\BibitemShut {NoStop}\bibitem [{\citenamefont {Kramers}(1930)}]{Kramers1930}\BibitemOpen
  \bibfield  {author} {\bibinfo {author} {\bibfnamefont {H.~A.}\ \bibnamefont
  {Kramers}},\ }\bibfield  {title} {\bibinfo {title} {Théorie générale de la
  rotation paramagnétique dans les cristaux},\ }\href
  {http://www.dwc.knaw.nl/DL/publications/PU00015981.pdf} {\bibfield  {journal}
  {\bibinfo  {journal} {Proceedings Koninklijke Akademie van Wetenschappen}\
  }\textbf {\bibinfo {volume} {33}},\ \bibinfo {pages} {959} (\bibinfo {year}
  {1930})}\BibitemShut {NoStop}\bibitem [{\citenamefont {Klein}(1952)}]{Klein1952}\BibitemOpen
  \bibfield  {author} {\bibinfo {author} {\bibfnamefont {M.~J.}\ \bibnamefont
  {Klein}},\ }\bibfield  {title} {\bibinfo {title} {{On a Degeneracy Theorem of
  Kramers}},\ }\href {https://doi.org/10.1119/1.1933118} {\bibfield  {journal}
  {\bibinfo  {journal} {American Journal of Physics}\ }\textbf {\bibinfo
  {volume} {20}},\ \bibinfo {pages} {65} (\bibinfo {year} {1952})}\BibitemShut
  {NoStop}\bibitem [{\citenamefont {Savit}(1980)}]{Savit1980}\BibitemOpen
  \bibfield  {author} {\bibinfo {author} {\bibfnamefont {R.}~\bibnamefont
  {Savit}},\ }\bibfield  {title} {\bibinfo {title} {Duality in field theory and
  statistical systems},\ }\href {https://doi.org/10.1103/revmodphys.52.453}
  {\bibfield  {journal} {\bibinfo  {journal} {Reviews of Modern Physics}\
  }\textbf {\bibinfo {volume} {52}},\ \bibinfo {pages} {453} (\bibinfo {year}
  {1980})}\BibitemShut {NoStop}\bibitem [{\citenamefont {Crapo}\ and\ \citenamefont
  {Whiteley}(1993)}]{Crapo1993}\BibitemOpen
  \bibfield  {author} {\bibinfo {author} {\bibfnamefont {H.}~\bibnamefont
  {Crapo}}\ and\ \bibinfo {author} {\bibfnamefont {W.}~\bibnamefont
  {Whiteley}},\ }\bibfield  {title} {\bibinfo {title} {Plane self stresses and
  projected polyhedra i: The basic pattem},\ }\href
  {http://hdl.handle.net/2099/1091} {\bibfield  {journal} {\bibinfo  {journal}
  {Structural Topology, 1993, núm. 20}\ }\textbf {\bibinfo {volume} {20}},\
  \bibinfo {pages} {55} (\bibinfo {year} {1993})}\BibitemShut {NoStop}\bibitem [{\citenamefont {Zhou}\ \emph {et~al.}(2019)\citenamefont {Zhou},
  \citenamefont {Zhang},\ and\ \citenamefont {Mao}}]{Zhou2019}\BibitemOpen
  \bibfield  {author} {\bibinfo {author} {\bibfnamefont {D.}~\bibnamefont
  {Zhou}}, \bibinfo {author} {\bibfnamefont {L.}~\bibnamefont {Zhang}},\ and\
  \bibinfo {author} {\bibfnamefont {X.}~\bibnamefont {Mao}},\ }\bibfield
  {title} {\bibinfo {title} {Topological boundary floppy modes in
  quasicrystals},\ }\href {https://doi.org/10.1103/physrevx.9.021054}
  {\bibfield  {journal} {\bibinfo  {journal} {Physical Review X}\ }\textbf
  {\bibinfo {volume} {9}},\ \bibinfo {pages} {021054} (\bibinfo {year}
  {2019})}\BibitemShut {NoStop}\bibitem [{\citenamefont {Senthil}(2004)}]{Senthil2004}\BibitemOpen
  \bibfield  {author} {\bibinfo {author} {\bibfnamefont {T.}~\bibnamefont
  {Senthil}},\ }\bibfield  {title} {\bibinfo {title} {Deconfined quantum
  critical points},\ }\href {https://doi.org/10.1126/science.1091806}
  {\bibfield  {journal} {\bibinfo  {journal} {Science}\ }\textbf {\bibinfo
  {volume} {303}},\ \bibinfo {pages} {1490} (\bibinfo {year}
  {2004})}\BibitemShut {NoStop}\bibitem [{\citenamefont {Zaanen}\ \emph {et~al.}(2015)\citenamefont {Zaanen},
  \citenamefont {Liu}, \citenamefont {Sun},\ and\ \citenamefont
  {Schalm}}]{Zaanen2015}\BibitemOpen
  \bibfield  {author} {\bibinfo {author} {\bibfnamefont {J.}~\bibnamefont
  {Zaanen}}, \bibinfo {author} {\bibfnamefont {Y.}~\bibnamefont {Liu}},
  \bibinfo {author} {\bibfnamefont {Y.-W.}\ \bibnamefont {Sun}},\ and\ \bibinfo
  {author} {\bibfnamefont {K.}~\bibnamefont {Schalm}},\ }\href
  {https://doi.org/10.1017/cbo9781139942492} {\emph {\bibinfo {title}
  {Holographic Duality in Condensed Matter Physics}}}\ (\bibinfo  {publisher}
  {Cambridge University Press},\ \bibinfo {year} {2015})\BibitemShut {NoStop}\bibitem [{\citenamefont {Senthil}\ \emph {et~al.}(2019)\citenamefont
  {Senthil}, \citenamefont {Son}, \citenamefont {Wang},\ and\ \citenamefont
  {Xu}}]{Senthil2019}\BibitemOpen
  \bibfield  {author} {\bibinfo {author} {\bibfnamefont {T.}~\bibnamefont
  {Senthil}}, \bibinfo {author} {\bibfnamefont {D.~T.}\ \bibnamefont {Son}},
  \bibinfo {author} {\bibfnamefont {C.}~\bibnamefont {Wang}},\ and\ \bibinfo
  {author} {\bibfnamefont {C.}~\bibnamefont {Xu}},\ }\bibfield  {title}
  {\bibinfo {title} {Duality between (2+1)d quantum critical points},\ }\href
  {https://doi.org/10.1016/j.physrep.2019.09.001} {\bibfield  {journal}
  {\bibinfo  {journal} {Physics Reports}\ }\textbf {\bibinfo {volume} {827}},\
  \bibinfo {pages} {1} (\bibinfo {year} {2019})}\BibitemShut {NoStop}\bibitem [{\citenamefont {Hull}\ and\ \citenamefont
  {Townsend}(1995)}]{Hull1995}\BibitemOpen
  \bibfield  {author} {\bibinfo {author} {\bibfnamefont {C.}~\bibnamefont
  {Hull}}\ and\ \bibinfo {author} {\bibfnamefont {P.}~\bibnamefont
  {Townsend}},\ }\bibfield  {title} {\bibinfo {title} {Unity of superstring
  dualities},\ }\href {https://doi.org/10.1016/0550-3213(94)00559-w} {\bibfield
   {journal} {\bibinfo  {journal} {Nuclear Physics B}\ }\textbf {\bibinfo
  {volume} {438}},\ \bibinfo {pages} {109} (\bibinfo {year}
  {1995})}\BibitemShut {NoStop}\bibitem [{\citenamefont {Maldacena}(1999)}]{Maldacena1999}\BibitemOpen
  \bibfield  {author} {\bibinfo {author} {\bibfnamefont {J.}~\bibnamefont
  {Maldacena}},\ }\bibfield  {title} {\bibinfo {title} {The large-n limit of
  superconformal field theories and supergravity},\ }\href
  {https://doi.org/10.1023/a:1026654312961} {\bibfield  {journal} {\bibinfo
  {journal} {International Journal of Theoretical Physics}\ }\textbf {\bibinfo
  {volume} {38}},\ \bibinfo {pages} {1113} (\bibinfo {year}
  {1999})}\BibitemShut {NoStop}\bibitem [{\citenamefont {Segal}(1968)}]{Segal1968}\BibitemOpen
  \bibfield  {author} {\bibinfo {author} {\bibfnamefont {G.}~\bibnamefont
  {Segal}},\ }\bibfield  {title} {\bibinfo {title} {Equivariant k-theory},\
  }\href {https://doi.org/10.1007/bf02684593} {\bibfield  {journal} {\bibinfo
  {journal} {Publications mathématiques de l'{IHÉS}}\ }\textbf {\bibinfo
  {volume} {34}},\ \bibinfo {pages} {129} (\bibinfo {year} {1968})}\BibitemShut
  {NoStop}\bibitem [{\citenamefont {Merkurjev}(2005)}]{Merkurjev2005}\BibitemOpen
  \bibfield  {author} {\bibinfo {author} {\bibfnamefont {A.~S.}\ \bibnamefont
  {Merkurjev}},\ }\bibfield  {title} {\bibinfo {title} {Equivariant k-theory},\
  }in\ \href {https://doi.org/10.1007/978-3-540-27855-9_18} {\emph {\bibinfo
  {booktitle} {Handbook of K-Theory}}}\ (\bibinfo  {publisher} {Springer Berlin
  Heidelberg},\ \bibinfo {year} {2005})\ pp.\ \bibinfo {pages}
  {925--954}\BibitemShut {NoStop}\bibitem [{\citenamefont {Constantin}\ and\ \citenamefont
  {Kolev}(2003)}]{Constantin2003}\BibitemOpen
  \bibfield  {author} {\bibinfo {author} {\bibfnamefont {A.}~\bibnamefont
  {Constantin}}\ and\ \bibinfo {author} {\bibfnamefont {B.}~\bibnamefont
  {Kolev}},\ }\bibfield  {title} {\bibinfo {title} {The theorem of
  {Kerékjártó} on periodic homeomorphisms of the disc and the sphere},\
  }\href {https://doi.org/10.5169/seals-61111} {\bibfield  {journal} {\bibinfo
  {journal} {L'Enseignement Mathématique}\ }\textbf {\bibinfo {volume} {40}},\
  \bibinfo {pages} {193} (\bibinfo {year} {2003})},\ \Eprint
  {https://arxiv.org/abs/math/0303256} {math/0303256} \BibitemShut {NoStop}\bibitem [{\citenamefont {Mazur}\ and\ \citenamefont {Ulam}(1932)}]{Mazur1932}\BibitemOpen
  \bibfield  {author} {\bibinfo {author} {\bibfnamefont {S.}~\bibnamefont
  {Mazur}}\ and\ \bibinfo {author} {\bibfnamefont {S.}~\bibnamefont {Ulam}},\
  }\bibfield  {title} {\bibinfo {title} {Sur les transformations
  isom{\'e}triques d’espaces vectoriels norm{\'e}s},\ }\href
  {https://doi.org/10.1080/00029890.2003.11920004} {\bibfield  {journal}
  {\bibinfo  {journal} {C. R. Acad. Sci. Paris}\ }\textbf {\bibinfo {volume}
  {194}},\ \bibinfo {pages} {116} (\bibinfo {year} {1932})}\BibitemShut
  {NoStop}\bibitem [{\citenamefont {Väisälä}(2003)}]{Vaisala2003}\BibitemOpen
  \bibfield  {author} {\bibinfo {author} {\bibfnamefont {J.}~\bibnamefont
  {Väisälä}},\ }\bibfield  {title} {\bibinfo {title} {{A Proof of the
  Mazur-Ulam Theorem}},\ }\href
  {https://doi.org/10.1080/00029890.2003.11920004} {\bibfield  {journal}
  {\bibinfo  {journal} {The American Mathematical Monthly}\ }\textbf {\bibinfo
  {volume} {110}},\ \bibinfo {pages} {633} (\bibinfo {year}
  {2003})}\BibitemShut {NoStop}\bibitem [{\citenamefont {Moln{\'{a}}r}(2015)}]{Molnar2015}\BibitemOpen
  \bibfield  {author} {\bibinfo {author} {\bibfnamefont {L.}~\bibnamefont
  {Moln{\'{a}}r}},\ }\bibfield  {title} {\bibinfo {title} {{General
  Mazur{\textendash}Ulam Type Theorems and Some Applications}},\ }in\ \href
  {https://doi.org/10.1007/978-3-319-18494-4_21} {\emph {\bibinfo {booktitle}
  {Operator Semigroups Meet Complex Analysis, Harmonic Analysis and
  Mathematical Physics}}}\ (\bibinfo  {publisher} {Springer International
  Publishing},\ \bibinfo {year} {2015})\ pp.\ \bibinfo {pages}
  {311--342}\BibitemShut {NoStop}\bibitem [{\citenamefont {Conway}\ and\ \citenamefont
  {Kochen}(2006)}]{Conway2006}\BibitemOpen
  \bibfield  {author} {\bibinfo {author} {\bibfnamefont {J.}~\bibnamefont
  {Conway}}\ and\ \bibinfo {author} {\bibfnamefont {S.}~\bibnamefont
  {Kochen}},\ }\bibfield  {title} {\bibinfo {title} {{The Free Will Theorem}},\
  }\href {https://doi.org/10.1007/s10701-006-9068-6} {\bibfield  {journal}
  {\bibinfo  {journal} {Foundations of Physics}\ }\textbf {\bibinfo {volume}
  {36}},\ \bibinfo {pages} {1441} (\bibinfo {year} {2006})}\BibitemShut
  {NoStop}\bibitem [{\citenamefont {Chiang}(2005)}]{Chiang2005}\BibitemOpen
  \bibfield  {author} {\bibinfo {author} {\bibfnamefont {T.}~\bibnamefont
  {Chiang}},\ }\bibfield  {title} {\bibinfo {title} {What's expected of us},\
  }\href {https://doi.org/10.1038/436150a} {\bibfield  {journal} {\bibinfo
  {journal} {Nature}\ }\textbf {\bibinfo {volume} {436}},\ \bibinfo {pages}
  {150} (\bibinfo {year} {2005})}\BibitemShut {NoStop}\bibitem [{\citenamefont {Aaronson}(2016)}]{Aaronson2016}\BibitemOpen
  \bibfield  {author} {\bibinfo {author} {\bibfnamefont {S.}~\bibnamefont
  {Aaronson}},\ }\bibfield  {title} {\bibinfo {title} {{The Ghost in the
  Quantum Turing Machine}},\ }in\ \href@noop {} {\emph {\bibinfo {booktitle}
  {The Once and Future Turing: Computing the World}}},\ \bibinfo {editor}
  {edited by\ \bibinfo {editor} {\bibfnamefont {S.~B.}\ \bibnamefont {Cooper}}\
  and\ \bibinfo {editor} {\bibfnamefont {A.}~\bibnamefont {Hodges}}}\ (\bibinfo
   {publisher} {Cambridge University Press},\ \bibinfo {year} {2016})\ \Eprint
  {https://arxiv.org/abs/1306.0159} {1306.0159} \BibitemShut {NoStop}\end{thebibliography}
\end{document}